\documentclass[prd,superscriptaddress,nofootinbib,floatfix]{revtex4}
\usepackage{bm,amsmath,amssymb,natbib}
\usepackage{ifthen}
\usepackage{aas_macros}
\usepackage{graphicx}

\usepackage[svgnames]{xcolor}

\newboolean{showcomments}
\setboolean{showcomments}{true}
\newcommand{\antcomment}[1]{\ifthenelse{\boolean{showcomments}}{{\color{DarkGreen} #1}}{}}
\newcommand{\riccomment}[1]{\ifthenelse{\boolean{showcomments}}{{\color{Maroon} #1}}{}}

\newcommand{\clh}{\mathcal{H}}
\newcommand{\la}{\langle}
\newcommand{\ra}{\rangle}
\newcommand{\Mpc}{\text{Mpc}}

\newcommand\del{\nabla}
\newcommand\grad{\bm{\nabla}}

\newcommand{\vv}{\mathbf{v}}
\newcommand{\vnhat}{\hat{\mathbf{n}}}
\newcommand{\vxhat}{\hat{\mathbf{x}}}
\newcommand{\vyhat}{\hat{\mathbf{y}}}
\newcommand{\vrhat}{\hat{\mathbf{r}}}

\newcommand{\vy}{\mathbf{y}}
\newcommand{\vx}{\mathbf{x}}
\newcommand{\vz}{\mathbf{z}}
\newcommand{\vk}{\mathbf{k}}
\newcommand{\vq}{\mathbf{q}}

\renewcommand\vr{\mathbf{r}}
\newcommand\vs{\mathbf{s}}

\newcommand\vkhat{\hat{\mathbf{k}}}

\newcommand{\xhat}{\hat{x}}
\newcommand{\yhat}{\hat{y}}
\newcommand{\rhat}{\hat{r}}
\newcommand\vsz{\mathsf{z}}
\newcommand\vsw{\mathsf{w}}
\newcommand\vst{\mathsf{t}}
\newcommand\vsu{\mathsf{u}}

\newcommand\mc{\mathsf{C}}
\newcommand\ma{\mathsf{A}_\phi}

\newcommand\trans[1]{#1^\mathsf{T}}

\newcommand\kpar{k_\parallel}%
\newcommand\qpar{q_\parallel}%
\newcommand{\rpar}{r_\parallel}
\newcommand{\kperp}{k_\perp}
\newcommand{\rperp}{r_\perp}

\newcommand{\dkp}[1][]{\ensuremath{\frac{d^3k_{#1}}{(2\pi)^{3}}}}%
\newcommand{\dqp}[1][]{\ensuremath{\frac{d^3q_{#1}}{(2\pi)^{3}}}}%
\newcommand{\presup}[1]{\ensuremath{^{\scriptscriptstyle #1}}}

\usepackage{hyperref}

\begin{document}

\title{Non-linear Redshift-Space Power Spectra}

\author{J. Richard Shaw}
\email{jrs65@ast.cam.ac.uk}
\author{Antony Lewis}
\homepage{http://cosmologist.info}

\affiliation{Institute of Astronomy, Madingley Road, Cambridge, CB3 0HA, UK.}

\date{\today}
\begin{abstract}
  Distances in cosmology are usually inferred from observed
  redshifts---an estimate that is dependent on the local peculiar
  motion---giving a distorted view of the three dimensional structure
  and affecting basic observables such as the correlation function and
  power spectrum.  We calculate the full non-linear redshift-space
  power spectrum for Gaussian fields, giving results for both the
  standard flat sky approximation and the directly-observable angular
  correlation function and angular power spectrum
  $C_l(z,z')$. Coupling between large and small scale modes boosts the
  power on small scales when the perturbations are small. On larger
  scales power is slightly suppressed by the velocities perturbations
  on smaller scales.  The analysis is general, but we comment
  specifically on the implications for future high-redshift
  observations, and show that the non-linear spectrum has
  significantly more complicated angular structure than in linear
  theory.  We comment on the implications for using the angular
  structure to separate cosmological and astrophysical components of
  21 cm observations.
\end{abstract}


\maketitle

\section{Introduction}

At the most fundamental level cosmological observations consist of
measurements of radiation intensity and frequency as a function of
angle on the sky. From these we can try to infer properties of the
Universe on our past light cone, and from them learn about
cosmology. To make more than the simplest inferences we must find a
reliable distance to the source we are observing. Fortunately if the
frequency of an emitting source is known, the redshift can be used as
a measurement of distance, allowing us to map our past light cone as a
function of angle and redshift. The observed redshift includes several
effects, but the most important is that from cosmological expansion
which allows us to estimate the distance. Secondary to this is the
doppler shifting from the peculiar velocity of the source along our
line of sight. For measurements of the displacement of a source from
us, the peculiar velocity quickly becomes negligible in comparison to
the cosmological redshifting. However, when measuring the separation
between spatially close points the correlated peculiar velocities can
have an important effect.  When inferring the statistics of
cosmological fluctuations it is therefore important to carefully model
the effect of velocities.

The universe is assumed to be spatially statistically homogeneous and
isotropic at a given time. The non-linear mapping between real space
(measured by comoving distance) and redshift space (measured by the
redshift $z$) means that a Gaussian field (with Gaussian densities and
velocities) will no longer be Gaussian when observed in redshift
space, and its power spectrum will also be different. In this paper we
show how to calculate the non-linear redshift-space power spectrum and
quantify the effects numerically. The linear result is
well-known~\cite{redshift:kaiser1987,Hamilton:1997zq}, but here we use
a non-perturbative approach to calculate results to all orders. As we
shall see, the non-linear corrections can be important at small scales
even at high redshift, and are therefore potentially important for
future high-redshift observations.

When the non-linear corrections become important, for full consistency
one should also calculate the non-linear evolution of the fields: an
initially Gaussian random field will be modified once non-linear
growth starts to be perturbatively
important~\cite{Makino92,Heavens:1998es,Scoccimarro:2004tg}. These
non-linear effects are more complicated to model, and depend on which
source is being observed; for example, the 21cm source evolution is
quite different to that of galaxy number counts. In this paper we
therefore neglect these complications, focussing on understanding the
important implications of the redshift-space mapping alone, with the
important caveat that our results must be generalized for application
to real observations. Our analysis is applicable to any observable
that can be reasonably approximated as having a Gaussian source field
with Gaussian velocities, and hence, within our approximation, applies
equally to biased source number counts or 21cm.

Since the line of sight defines a vector field on the past light cone,
the light cone as a function of redshift and angle is only
statistically isotropic about the centre of symmetry --- the
observation point. The inferred angular structure of the field about
other points therefore gives information about the local velocity
field. In linear theory the velocities are simply related to the total
density when dark matter and baryon velocities are the same. Hence an
observation of the velocities could be used to constrain directly the
cosmological density field independently of the sources, which could
be hard to model because of complicated astrophysics. We show that the
non-linear corrections to the angular structure can be important when
attempting to measure the densities this way.

This paper will continue as follows: In the next sub-section
(Section~\ref{sec:lineartheory}) we briefly overview the results from
linear theory. In Section~\ref{sec:nonlinear} we introduce our method
for calculating the non-linear redshift-space power
spectra. Section~\ref{sec:radiative} discusses the differences
encountered when calculating the power spectrum of radiative fields
like the brightness compared to spatial densities such as the matter
perturbation. In Section~\ref{sec:flatsky} we calculate the
three-dimensional power spectrum and discuss the results. To go beyond
this to the full-sky, in Section~\ref{sec:curvedsky}, we calculate the
angular correlation function and angular power spectrum. Finally we
discuss what bearing our results have on high-redshift 21cm
observations in Section~\ref{sec:21cm}.

Throughout the rest of this paper we assume a standard flat
concordance $\Lambda$CDM cosmology with matter densities $\Omega_c h^2
= 0.104$, $\Omega_b h^2 = 0.022$ for dark and baryonic matter
respectively. We take a Hubble parameter of $H_0 = 73
\mathrm{km\,s}^{-1} \mathrm{Mpc}^{-1}$, and optical depth to Thomson
scattering $\tau = 0.09$. We use a primoridal power spectrum with
constant spectral index $n_s = 0.95$ and amplitude $A_s = 2.04 \times
10^{-9}$ at a scale of $0.05 \mathrm{Mpc}^{-1}$. Furthermore we
neglect the neutrino masses which should have small effects at high
$k$.

\subsection{Redshift-space mapping and linear result}
\label{sec:lineartheory}
Assuming the redshift is entirely cosmological, the comoving distance
to an object at redshift $z$ is
\begin{equation}
  \chi_z = \int^z_0\frac{dz'}{(1+z')\mathcal{H}(z')},
\end{equation}
where $\mathcal{H}$ is the comoving Hubble parameter and throughout we
use natural units with $c=1$.
In general this equation \emph{defines} what we call the
redshift-space distance, which can easily be calculated from the
observed redshift given a background cosmology.  However it is not
equal to the actual comoving distance in a perturbed universe: the
peculiar velocity means that the actual comoving distance $\chi$ at
redshift $z$ is not $\chi_z$, but also depends on the local velocity
field $\vv(\vx)$. Neglecting local evolution of the background, small
lensing and general-relativistic effects, and assuming that the
peculiar velocities are non-relativistic, the comoving distance is in
fact
\begin{equation}
  \chi = \chi_z - \left.\vv(\vx) \cdot \vnhat / \mathcal{H}\right|_z\; .
\end{equation}
Note that we assume the peculiar velocity of the observer is removed
from the observed redshifts so that only the source velocity matters.
From here onwards we write $\phi(\vx)\equiv \vv(\vx) \cdot \vnhat /
\mathcal{H} $, and denote our coordinates in real space as $\vx = \chi
\vnhat$, and redshift space as $\vs = \chi_z \vnhat$, such that the mapping
between the two is
\begin{equation}
\vs = \vx + \phi(\vx) \vnhat \; .
\end{equation}


The effect at first order in the power spectrum is well known and easy
to calculate \cite{redshift:kaiser1987}. Transforming the mass in a
small volume element from real to redshift space using the Jacobian
factor we have
\begin{equation}
d^3s = d^3x \left\lvert\frac{\partial \vs}{\partial \vx}\right\rvert \; .
\end{equation}
In the distant observer approximation we neglect the curvature of the
sky, and thus the Jacobian factor contains only the line of sight term
$\frac{\partial s}{\partial \chi} = 1 + \phi'$, with the prime
denoting differentiation with respect to the line of sight
direction. We discuss this point in more depth in
Section~\ref{sec:radiative}. Conserving the mass in the elements gives
\begin{equation}
\overline{\rho}[1 + \Delta_s(\vs)]\, d^3s = \overline{\rho}[1 + \Delta(\vx)]\, d^3x
\end{equation}
and hence
\begin{equation}
\label{eq:jac_trans}
 \Delta_s(\vs) = \frac{\Delta(\vx)-\phi'(\vx)}{1 + \phi'(\vx)},
\end{equation}
where the source perturbation in real space is $\Delta$ and in
redshift space is $\Delta_s$. Expanding this to first order gives the
redshift-space perturbation
\begin{equation}
\label{eq:first_source}
  \Delta_s(\vs) \approx \Delta(\vs) - \phi'(\vs) \; .
\end{equation}
Note that in this we use the fact that $\vs = \vx$ at first order to
transform the arguments. In Fourier space we have
\begin{equation}
\Delta_s(\vk) = \Delta(\vk) - i \kpar \phi(\vk),
\end{equation}
where $\kpar \equiv \vnhat \cdot \vk$.  The quantity we are interested
in is the power spectrum $P_s$ of $\Delta_s$ given by
\begin{equation}
\label{eq:ps_1}
P_s(\vk) = P_\Delta(k) + 2 i \kpar P_{\Delta\phi}(\vk) + \kpar^2 P_\phi(\vk),
\end{equation}
where $P_\Delta$, $P_{\Delta\phi}$ and $P_\phi$ are generated by the
obvious contraction of $\Delta$ and $\phi$. For irrotational flows we
can link the velocity vector field to an underlying scalar
perturbation $\delta_v$, defined by the relation
\begin{equation}
\label{eq:def_delv}
\del\cdot\vv(\vx) = - \mathcal{H}\, \delta_v(\vx) \; .
\end{equation}
This definition applies generally and makes no constraints on our
tracer $\Delta$.  It is, however, motivated by the continuity equation
for pressureless matter in the linear growth era. In this regime the
scalar perturbation to the velocities $\delta_v$ simply relates to the
total matter perturbation $\delta_m$ via $\delta_v = f \delta_m$,
where $f$ is the derivative of the linear growth factor for matter
perturbations, $f \equiv d\ln{D_+} / d\ln{a}$. The equivalent Fourier
space definition to Eq.~\eqref{eq:def_delv} is $\vv(\vk) = i
\mathcal{H} \left(\vk / k^2\right)\, \delta_v(\vk)$, so, writing
$\mu_k = \vnhat \cdot \vkhat = \kpar / k$, Eq.~\eqref{eq:ps_1} becomes
\begin{equation}
\label{eq:pslin}
  P_{s}(\vk) = P_\Delta(k) + 2 \mu_k^2 P_{\Delta v}(\vk) + \mu_k^4 P_v(\vk).
\end{equation}
If the field we consider is a linearly biased tracer of the underlying
matter distribution, such as a simplistic model of galaxy number
counts, we would have $\Delta = b \delta_m$, with $b$ the linear bias
factor. Assuming no velocity bias this gives the classic Kaiser result
(see \cite{redshift:kaiser1987})
\begin{equation}
  P_{g,s}(\vk) = b^2 \bigl(1 + b^{-1} f \mu_k^2\bigr)^2\, P_\delta(k) \; .
\end{equation}


\section{Non-linear Power Spectrum}
\label{sec:nonlinear}
\noindent The contributions to the redshift-space power spectrum beyond
linear theory could be calculated by a perturbative expansion. We
discuss perturbative relationships between real and redshift space in
Appendices~\ref{app:perturb} and~\ref{app:series}. However as we might
expect this approach becomes tedious above second order, and features
independently large terms that nearly exactly cancel. The reason for
this behaviour is that the effective displacement caused by a velocity
becomes larger than the perturbation wavelength on small scales, so
the small scale contribution to $\Delta(\vs)$ is very different from
$\Delta(\vx)$. However most of this displacement comes from the
coherent large-scale velocity field, which has little effect on the
\emph{difference} of the velocities that is important for an
observable change in the correlation function. A bulk radial
displacement is not observable in the flat-sky approximation. For this
reason an approach based on transforming the real-space correlation
functions may be significantly better. This is the approach we adopt
here, which allows us to calculate a simple non-perturbative result
for the redshift-space power spectrum.

We would like to find how a Gaussian density field $\Delta(\vx)$
appears in redshift space. Our starting point is from the conservation
of field mass in a small volume element, in both real-space and
redshift space
\begin{equation}
[1 + \Delta_s(\vs)]\, d^3s = [1 + \Delta(\vx)]\, d^3x \; .
\end{equation}
We emphasize that this is for a density
field such as source counts, e.g. the galactic number
density. Radiative fields such as the brightness and brightness
temperature are different since the measurement is then of observed photon counts, rather than source number counts;
we address this is
Section~\ref{sec:radiative}. With this restriction in mind we multiply
both sides by $e^{-i\vk\cdot\vs}$ and integrate, finding that
\begin{equation}
\label{eq:fint1}
\int [1 + \Delta_s(\vs)]\, e^{-i\vk\cdot\vs}\,d^3s = \int [1 +
  \Delta(\vx)]\, e^{-i\vk\cdot\vs}\,d^3x \; ,
\end{equation}
and substituting $\vs = \vx+ \vnhat_x \phi(\vx)$ we then have
\begin{equation}
(2\pi)^3\delta^3(\vk) + \Delta_s(\vk) = \int d^3x \: e^{-i \vk \cdot \vx} [1 + \Delta(\vx)]\, e^{-i \kpar \phi(\vx)} \; ,
\label{eq:deltak_delta}
\end{equation}
where $\delta^3(\vk)$ is the Dirac delta-function that we can neglect
provided we limit ourselves to the behaviour at $\vk \ne 0$. To
calculate the power spectrum we use
\begin{equation}
\label{eq:kq_noneval}
\bigl\langle\Delta_s(\vk)\,\Delta_s(\vq)\bigr\rangle = \iint d^3x\,d^3y\; e^{-i[\vk \cdot \vx + \vq \cdot \vy]}\, \Bigl\langle[1 + \Delta(\vx)]\,[1 + \Delta(\vy)] \,e^{-i[\kpar \phi(\vx) + \qpar \phi(\vy)]}\Bigr\rangle \; ,
\end{equation}
where $\qpar = \vq \cdot \vnhat_y$, and $\vnhat_y = \vy / y$. To
calculate the expectation value we assume that all the fields are
Gaussian. Writing the fields as a vector $\trans{\vsz} =
\left(\Delta(\vx), \Delta(\vy), \phi(\vx), \phi(\vy)\right)$, and
defining a further vector $\trans{\vsw} = -i\left(0, 0, \kpar,
  \qpar\right)$, we calculate the expectation values $\langle
e^{\trans{\vsw}\vsz}\rangle$, $\langle\vsz \,e^{\trans{\vsw}
  \vsz}\rangle$ and $\langle\vsz \,\trans{\vsz} e^{\trans{\vsw}
  \vsz}\rangle$, defined by
\begin{equation}
\Bigl\langle (\ldots)\: e^{\trans{\vsw} \vsz}\Bigr\rangle = \frac{1}{(2\pi)^2 \det^{1/2}\mc}\int d^4z \, \exp{\left[-\frac{1}{2} \trans{\vsz}\mc^{-1}\vsz + \vsw\cdot\vsz\right]}\: (\ldots) \; ,
\end{equation}
where the $\mc$ is the covariance matrix of the fields $\mc =
\left\langle\vsz\,\trans{\vsz}\right\rangle$. We complete the square
in the Gaussian integral to evaluate it, giving
\begin{subequations}
\label{eq:exp_val}%
\begin{equation}
\label{eq:exp_scalar}
\Bigl\langle e^{\trans{\vsw}\vsz}\Bigr\rangle = e^{\frac{1}{2}\trans{\vsw}\mc\vsw} .
\end{equation}
To calculate the remaining two expectation values we take the partial
derivatives with respect to $\vsw$:
\begin{align}
\label{eq:exp_vector}
\Bigl\langle\vsz \,e^{\trans{\vsw}\vsz}\Bigr\rangle & = e^{\frac{1}{2}\trans{\vsw}\mc\vsw} \, \mc\vsw \; ,\\ \label{eq:exp_tensor}
\Bigl\langle\vsz \,\trans{\vsz} e^{\trans{\vsw}\vsz}\Bigr\rangle & = e^{\frac{1}{2}\trans{\vsw}\mc\vsw} \left[\mc + \mc\,\vsw\,\trans{\vsw}\,\mc\right] \; .
\end{align}
\end{subequations}
The results of Eq.~\eqref{eq:exp_val} allow us to evaluate
Eq. \eqref{eq:kq_noneval}: we take \eqref{eq:exp_scalar}, the 1 and 2
components of \eqref{eq:exp_vector}, corresponding to $\Delta(\vx)$
and $\Delta(\vy)$, and the 1,2 component of \eqref{eq:exp_tensor},
from $\Delta(\vx)\Delta(\vy)$, and sum them to construct the
expectation value of Eq.~\eqref{eq:kq_noneval}. The required
components are
\begin{subequations}
\begin{align}
\trans{\vsw}\mc\vsw & = -\kpar^2 C_{\phi}(\vx,\vx)  - \qpar^2  C_\phi(\vy,\vy) - 2 \kpar\qpar C_\phi(\vx,\vy) \; ,\\
[\mc \cdot \vsw]_1 + [\mc \cdot \vsw]_2 & = -i\bigl[\qpar C_{\Delta\phi}(\vx,\vy) + \kpar  C_{\Delta\phi}(\vy,\vx)\bigr] \; ,\\
[\mc + \mc \vsw \trans{\vsw} \mc]_{12} & = C_\Delta(\vx,\vy) - \kpar \qpar C_{\Delta\phi}(\vx,\vy)C_{\Delta\phi}(\vy,\vx) \; ,
\end{align}
\end{subequations}
where we have defined $C_{ab}(\vx,\vy) = \left\langle a(\vx) b(\vy)
\right\rangle$. Note that statistical isotropy of the underlying
correlation requires $\langle \Delta(\vx) \vv(\vx)\rangle = 0$ and
hence the definition of the $\phi$ field means that
$C_{\Delta\phi}(\vx, \vx) = 0$. Combining the above, the expectation
value $\left\langle\Delta_s(\vk)\,\Delta_s(\vq)\right\rangle$
evaluates to
\begin{multline}
\label{eq:dkdq}
\bigl\langle\Delta_s(\vk)\,\Delta_s(\vq)\bigr\rangle = \iint d^3x\,d^3y\; e^{-i \left[\vk \cdot \vx + \vq \cdot \vy\right] } e^{-\frac{1}{2} \left[\kpar^2 C_{\phi}(\vx,\vx)  + \qpar^2  C_\phi(\vy,\vy) + 2 \kpar\qpar C_\phi(\vx,\vy)\right]} \\
 \times \left[1 + C_{\Delta}(\vx,\vy) - i \qpar C_{\Delta\phi}(\vx,\vy) - i \kpar C_{\Delta\phi}(\vy,\vx) - \kpar \qpar C_{\Delta\phi}(\vx,\vy) C_{\Delta\phi}(\vy,\vx)\right] \; .
\end{multline}
This result can now be used to calculate the flat-sky power spectrum
$P(\vk)$ and the directly-observable angular power spectrum
$C_l(z,z')$, as we show in the following sections.

It is possible to extend this method to calculation of higher n-point
functions, such as the bi-spectrum and higher moments, allowing
investigation of the non-Gaussianity introduced solely by the
redshift-space distortions. This is conceptually simple, we simply
take further moments of Eq.~\eqref{eq:deltak_delta} giving
\begin{equation}
\la \Delta_s(\vk_1)\Delta_s(\vk_2)\dotsm \Delta_s(\vk_n)\ra = \int \left(\prod_{j=1}^n d^3 x_j e^{-i[\vk_j\cdot\vx_j]}\right) \left\la\prod_{i=1}^n  [1 + \Delta(\vx_i)] e^{-i[{\kpar}_i\phi(\vx_i)]} \right\ra \; ,
\end{equation}
where we have continued to neglect the behaviour at $\vk = 0$. This
can be evaluated in the same manner as above, though that is beyond
the scope of this paper, we will limit ourselves to the power
spectrum.


\section{Radiative Fields and the Distant Observer Approximation}
\label{sec:radiative}
\noindent Both the matter density field, and galactic number density are
examples of spatial densities where the conserved quantity we consider
in the transformation between real and redshift space is the mass in a
small volume element
\begin{equation}
\rho_s(\vs) \, d^3s = \rho(\vx) \, d^3x \; .
\end{equation}
This was the line we proceeded along in the previous section. However
for radiative quantities such as the brightness we have a subtly
different result: if we radially displace a number of sources we still
observe the same number, however the brightness is less because we
receive fewer photons from a source that is more distant.  For a
detector of area $dA$, receiving frequencies in a range $d\nu$ about
$\nu$ from a source region of solid angle $d\Omega$, the brightness
$I_\nu$ is defined by the energy received $dE$ in a short time $dt$
\begin{equation}
dE = I_\nu\, dA\, d\Omega\, d\nu\, dt \;,
\end{equation}
or simply the brightness $I_\nu$ is the flux onto a detector at a
frequency $\nu$ from a source per unit solid angle per unit frequency.
For radiative fields the fundamental observed quantity  is
$I_\nu\, d\Omega\, d\nu$, the flux in a frequency range $\nu$ to $\nu
+ d\nu$, from a solid angle $d\Omega$. The redshift is determined by
the shift from the source frequency $\nu_0$, and thus the frequency
interval $d\nu$ gives the radial distance interval in real or redshift
space. The conservation equation, neglecting factors of $\mathcal{H}$,
is then
\begin{equation}
\label{dInu}
I_\nu(\vs)\, d\Omega ds = I_\nu(\vx)\, d\Omega dx \; .
\end{equation}
where the subtle distinction between $I_\nu(\vs)$ and $I_\nu(\vx)$ is
that in the latter we remove the distortion of the frequency interval
$d\nu$ caused by the peculiar motion.  In the Rayleigh-Jeans
approximation (excellent for typical 21cm line observation) the
brightness temperature is $T_b(\nu) \approx I_\nu c^2 / 2 k_b \nu^2$,
so this result also holds for the brightness temperature.  Using $s =
x+ \phi(\vx)$ this implies that
\begin{equation}
\label{eq:jac_radiative}
 \Delta_{s,T_b}(\vs) = \frac{\Delta_{T_b}(\vx)-\phi'(\vx)}{1 + \phi'(\vx)},
\end{equation}
which was only an approximation in the case of number counts,
Eq.~\eqref{eq:jac_trans}. We discuss the perturbative expansion of
this result in Appendix~\ref{app:perturb}. To follow the number count
derivation we must take the Fourier transform, and so convert the
small parameter space region $d\Omega ds$ into the small volume $d^3s
= s^2 d\Omega ds$ (similarly for real space), and hence write
Eq.~\eqref{dInu} as
\begin{equation}
[1 + \Delta_{s,T_b}(\vs)] \, d^3s = [1 + \Delta_{T_b}(\vx) ] \left(1 + \frac{\phi(\vx)}{x}\right)^2 \, d^3x \; .
\end{equation}
If we simply follow through the analysis of
Section~\ref{sec:nonlinear} we come unstuck because of the $1 +
\phi/x$ term, which would make the analysis significantly more
complicated (though not intractable).  The simplifying solution is to
apply an approximation that is not required in the spatial density
case, the distant observer approximation. Given that we are observing
at large distances relative to the velocity displacement $\phi$, and
that the distortions are sourced largely by the gradients of the
velocity field, we assert that for all scales of interest $\phi(\vx) /
x \ll \phi'(\vx)$ and set $1 + \phi/x \approx 1$. At high redshift ($z
> 5$) we find $\phi_\text{rms} / x$ to be at most of order $10^{-3}$
whilst $\phi'_\text{rms}$ is consistently of order $1$, so we expect
this to be a reasonable approximation. After this it is possible to
apply all the previous analysis to radiative fields such as
$\Delta_{T_b}$ as well as density fields.

Applying the distant observer approximation not only allows us to consider
radiative fields, but allows a simplification of the preceding
analysis in all cases. Starting from Eq.~\eqref{eq:fint1} we transform
the $\delta$-function generating term on the LHS by substituting in
explicitly for $\vx$ and writing it as
\begin{equation}
\label{eq:del_sub}
\int e^{-i\vk\cdot\vs}\,d^3s = \int e^{-i \vk\cdot\left[\vx + \vnhat \phi(\vx)\right]} \left(1+\frac{\phi(\vx)}{x}\right)^2 \left(1 + \phi'(\vx)\right) d^3x.
\end{equation}
Invoking the distant observer approximation removes the $\phi/x$
term, and canceling off the lowest order terms on both sides leaves us
with
\begin{equation}
\label{Deltas_21cm}
\Delta_s(\vk) = \int d^3x \: e^{-i \vk \cdot \vx} \bigl[\Delta(\vx) - \phi'(\vx)\bigr] e^{-i \kpar \phi(\vx)} \; .
\end{equation}
Note that this holds for all $\vk$ including $\vk = 0$ unlike the
previous formulation. For number counts this approximation neglects
first order $\phi/x$ terms, but for radiative fields it is actually
correct to first order, neglecting only terms ${\cal O}(\Delta\phi/x)$
and higher. Conceptually this is because if you radially displace a
volume at redshift $z$ in an angle $d\Omega$ the physical volume
corresponding to that $d\Omega$ increases $\propto r^2$, giving a
linear ${\cal O}(\phi/x)$ change to the number of sources
(c.f. Ref.~\cite{Papai:2008bd}). However by the inverse-square law the
fraction of photons received from each source goes down by $1/r^2$, so
the number of photons received is invariant at first order. To ensure
that the result for number counts contains all the effects at first
order we simply preserve the linear $\phi / x$ term in
Eq.~\eqref{eq:del_sub}.

Comparison with Eq.~\eqref{eq:jac_radiative} shows that the quantity
$\Delta(\vx) - \phi'(\vx)$ is the source of redshift distortions at
first order. Writing the redshift-space perturbation in this form
makes it clear where the contributions are coming from, and more
obvious how it reduces to the first order result. Given its importance
we will denote the first order source as $\alpha(\vx) = \Delta(\vx) -
\phi'(\vx)$ from now on. To progress towards the power spectrum we
follow the same lines as Eq.~\eqref{eq:kq_noneval} to
Eq.~\eqref{eq:dkdq} with the only change that we average over
$\trans{\vsz} = \left(\alpha(\vx), \alpha(\vy), \phi(\vx),
  \phi(\vy)\right)$ to calculate the expectation. Finally we have the
$\left\langle\Delta_s(\vk)\,\Delta_s(\vq)\right\rangle$ in the distant
observer approximation
\begin{multline}
\bigl\langle\Delta_s(\vk)\,\Delta_s(\vq)\bigr\rangle = \iint d^3x\,d^3y\; e^{-i \left[\vk \cdot \vx + \vq \cdot \vy\right] } \exp{\left(-\frac{1}{2} \left[\kpar^2 C_{\phi}(\vx,\vx)  + \qpar^2  C_\phi(\vy,\vy) + 2 \kpar\qpar C_\phi(\vx,\vy)\right]\right)} \\
 \times \left[C_{\alpha}(\vx,\vy) - \kpar \qpar C_{\alpha\phi}(\vx,\vy) C_{\alpha\phi}(\vy,\vx)\right] \; .
\end{multline}
To keep this correct for spatial densities at first order we must use
$\alpha(\vx) = \Delta(\vx) - \phi'(\vx) - 2\phi(\vx) / x$, giving the
linear result without having assumed the distant observer
approximation.

\section{Power Spectra on the Flat-Sky}
\label{sec:flatsky}
\noindent We first consider the flat-sky approximation, appropriate
for a small patch of sky sufficiently thin in redshift that evolution
along the light cone can be neglected. The patch is assumed to be at a
large distance and subtending a small angle so that $\vnhat \approx
\vnhat'$ across the patch. Since we are neglecting evolution, in a
statistically homogenous universe with isotropy broken locally only by
the line of sight direction the correlation functions should be a
function of $r = \lvert\vx - \vy\rvert$ and $\mu_r \equiv \vnhat\cdot
\hat{\vr}$ only, so
\begin{subequations}
\begin{align}
C_\Delta(\vx,\vy) &= \xi_{\Delta}(r)  \\
C_{\Delta\phi}(\vx,\vy) &= \xi_{\Delta\phi}(r, \mu_r)  \\
C_\phi(\vx,\vy) &= \xi_\phi(r, \mu_r) \; .
\end{align}
\end{subequations}
Changing one integration variable from $\vx$ to $\vr$ in
Eq.~\eqref{eq:dkdq}, we can then perform the integration over $\vy$
using
\begin{equation}
\int d^3y\: e^{-i \vy \cdot \left(\vk + \vq\right)} = (2 \pi)^3 \delta^3(\vk + \vq) \; .
\end{equation}
By definition the power spectrum is
\begin{equation}
\bigl\langle\Delta_s(\vk)\,\Delta_s(\vq)\bigr\rangle = (2 \pi)^3 \delta^3(\vk + \vq)\, P_s(\vk),
\end{equation}
and hence we identify $P_s(\vk)$ as
\begin{equation}
P_s(\vk) = \int d^3r \: e^{-i\vk \cdot \vr}\, \left[1 + \xi_{\Delta}(r) + 2 i \kpar \xi_{\Delta\phi}(r, \mu_r) - \kpar^2 \xi_{\Delta\phi}(r, \mu_r)^2 \right]e^{-\kpar^2 \left[\xi_{\phi}(0) - \xi_{\phi}(r, \mu_r)\right]} \; .
\label{eq:flatsky_final3d}
\end{equation}
The correlation functions above are dependent only on the angle
between $\vr$ and $\vnhat$, and hence are azimuthally symmetric,
allowing us to integrate out this dependence. If we separate the
exponential term as $e^{-i\vk \cdot \vr} = e^{-i\kpar \rpar}\,
e^{-i\kperp \rperp \cos{\varphi}}$ we can integrate over $\varphi$,
and use the identity:
\begin{equation}
\frac{1}{2\pi} \int_0^{2\pi} \!\exp{(-i\,x \cos{\varphi})}\,d\varphi \equiv J_0(x),
\end{equation}
where $J_0(x)$ is the zeroth Bessel function of the first
kind. Furthermore knowing that the result will be real, we can write
separate real and imaginary parts into the cosine and sine parts of
the exponential. Combining these we have
\begin{align}
\label{eq:flatsky_final}
P_s(\vk) = \,\,& 4\pi \int^{\infty}_0\! dr \int^1_{0}\! d\mu_r\: r^2 J_0(\kperp r \sqrt{1-\mu_r^2})\, e^{-\kpar^2 \left[\xi_{\phi}(0) - \xi_{\phi}(r, \mu_r)\right]} \notag \\
& \times \biggl[\cos{(\kpar r \mu_r)} \Bigl[1 + \xi_{\Delta}(r) - \kpar^2 \xi_{\Delta\phi}(r, \mu_r)^2 \Bigr] + 2 \kpar \sin{(\kpar r \mu_r)}\, \xi_{\Delta\phi}(r, \mu_r)\biggr] \; .
\end{align}
This is the final form, suitable for numerical
evaluation. Unfortunately the integral is highly oscillatory, but we
must still include the structure in the integrand across a large range
between $k\: \Mpc \approx 10^{-3}$--$10^3$. This includes a very large
number of oscillation and thus requires careful
evaluation. Calculation of the correlation functions $\xi_\Delta$,
$\xi_{\Delta\phi}$ and $\xi_\phi$ from the relevant power spectra is
considered in Appendix \ref{app:correlation}.

A result equivalent to Eq.~\eqref{eq:flatsky_final} has been derived
previously in Ref.~\cite{Scoccimarro:2004tg}, but numerical
calculation was not attempted because the focus was on low redshifts
where other non-linear effects are very important. Here we calculate
the effects at high redshift, discuss the physical origin of the
various effects, and in Section~\ref{sec:curvedsky} also generalize to
the directly observable angular power spectrum. We also note from
Section~\ref{sec:radiative} that Eq.~\eqref{eq:flatsky_final} can
alternatively be written in terms of the correlation functions of the
first-order source $\alpha(\vx)$ as
\begin{equation}
\label{eq:flatsky_final2}
P_s(\vk) = \,\, 4\pi \int^{\infty}_0\! dr \int^1_{0}\! d\mu_r\: r^2 J_0(\kperp r \sqrt{1-\mu_r^2})\,
\cos{(\kpar r \mu_r)}
  \Bigl[\xi_{\alpha}(r, \mu_r) - \kpar^2 \xi_{\alpha\phi}(r, \mu_r)^2 \Bigr]e^{-\kpar^2 \left[\xi_{\phi}(0) - \xi_{\phi}(r, \mu_r)\right]} \; .
  \end{equation}

\begin{figure}
\begin{center}
\includegraphics[width=0.8\textwidth]{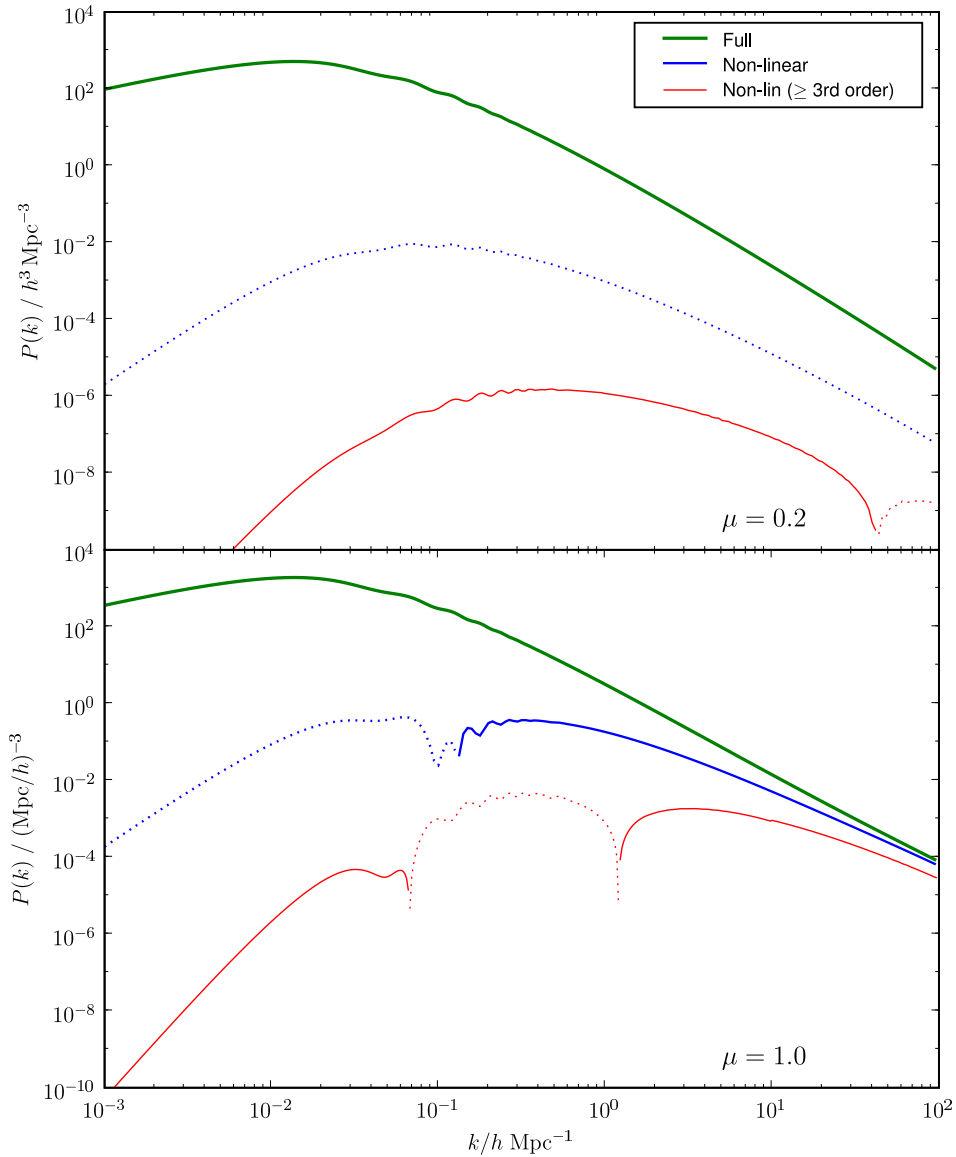}
\end{center}
\caption{The full dark matter power spectrum, and the non-linear
  contributions at redshift $z = 10$. We plot two values of $\mu$, a
  small value $\mu = 0.2$ in the upper plot and the completely
  parallel case $\mu = 1.0$ in the lower plot. The solid lines are for
  positive values, the dotted lines are negative. Whilst the
  non-linear contributions are negative for $\mu = 0.2$, the
  contributions at higher $\mu$ actually boost the power on small
  scales. We also plot the non-linear contributions greater than
  second order in the power spectrum, this shows that second order
  perturbation theory is largely inadequate at high-$k$ and high
  $\mu$, at about 5\% at $k=10 h \, \mathrm{Mpc}^{-1}$.}
\label{fig:comp_mu}
\end{figure}

In Figure \ref{fig:comp_mu} we compare the non-linear results at
redshift 10 for two distinct values of $\mu_k$. There are two distinct
effects taking place here: firstly at low $\mu_k$ there is a
suppression of power across all scales; secondly at high $\mu_k$ there
is an increase in power which overcomes the general suppression at
large values of $k$. The effect reaches the 1\% level at around $k =
0.3 h\, \mathrm{Mpc}^{-1}$. If we calculate the rms perturbation in
spheres of half of this wavelength $\pi / k \approx 10.5 h^{-1}
\mathrm{Mpc}$, we find $\sigma_{10.5} \approx 0.077$. Thus at this
scale perturbations are still firmly linear, and this effect should be
significant relative to any non-linear evolution.

We can gain some insight into the physical origin of these effects by
considering the leading order perturbative corrections, that is those
second order in the power spectra. We make use of some of the results
from Appendix \ref{app:series} where we examine the perturbative
expansion and second order asymptotics.

The general suppression can be understood from the form of the
perturbative result at large scales. Taking the result from
\eqref{eq:largescale}, we find that on large scales the non-linear
contribution ($\Delta P_s(\vk) = P_s(\vk) - P_s^{\mathrm{lin}}(\vk)$)
for fully correlated fields is
\begin{equation}
\Delta P_s(\vk) \sim -\kpar^2 \xi_\phi(0) P_s^{\text{lin}}(\vk)  \; .
\end{equation}
To gain insight into this note that $\xi_\phi(0)$ is the point
line-of-sight velocity variance in Hubble units, which serves to wash
out a large-scale mode with wavenumber $\vk$ in the line-of-sight
direction by a fraction ${\cal O}(\kpar\xi_\phi(0)^{1/2})$ of a
wavelength. This leads to a suppression of large-scale power.


The expansion of the perturbative result for large $k$ suggests a
source of the small-scale boost in power: the superposition of
large-scale modes on top of modes at that $k$. The contributions in
Eq.~\eqref{eq:smallscale} are complicated, though schematically they
are of the form
\begin{equation}
\Delta P_s(\vk)  \sim P_{\phi'}(\vk) \xi_\alpha(0) + P_{\alpha\phi'}(\vk) \xi_{\alpha\phi'}(0) + P_{\alpha}(\vk)  \xi_{\phi'}(0) \; ,
\end{equation}
where we have neglected constants and angular dependence, and have
approximated $k \frac{d P_a(\vk)}{dk} \approx \mathrm{const.}\times
P_a(\vk)$ which is good for large $k$ in the tail of the spectrum. All
terms are of the form power spectrum at some $\vk$ times the point
variance of another quantity from larger scales. The first term (which
is essentially exact) represents the superposition of velocity
gradients on the point redshift-space power coming from larger
scales. The other terms are similar, but contain complicated angular
behaviour which we have omitted.

At lower redshift, when terms above second order become important, the
exponent term in Eq.~\eqref{eq:flatsky_final} becomes large unless
$\vr \sim 0$. This leads to an exponential suppression of the coupling
from larger scales, reflecting the fact that once small scale
velocities effectively wipe out the power by line-of-sight smearing,
this wins over the boost due to superimposing larger-scale modes.  The
calculation is of course not reliable in this regime due to
significant non-Gaussianity and non-linear evolution, nonetheless the
qualitative effect is well known as the Fingers of God, when
non-linear clusters contribute significant small-scale
velocities~\cite{paper:jackson1972}. An extra uncorrelated Gaussian
point velocity variance $\sigma_v^2$ can easily be included in our
model by making the substitution $\xi_\phi(0) \rightarrow \xi_\phi(0)
+ \sigma_v^2/3\clh^2$. This has the effect that $P_s(\vk) \rightarrow
e^{-\kpar^2\sigma_v^2/3\clh^2} P_s(\vk)$, so that power on scales
smaller than the redshift-space spread are exponentially
suppressed. This describes the effect of finite line width due to the
local thermal motion when considering diffuse 21cm, and also roughly
the effect of non-linear virial motion within clusters when measuring
number count power spectra on much larger scales. For further
discussion of an approximate effective model at low redshift when
non-linear evolution is important see
Ref.~\cite{Scoccimarro:2004tg,Percival:2008sh}.

In Figure~\ref{fig:comp_mu} we also plot the contributions to the
power spectra with the terms at first and second order in the linear
spectrum subtracted off, showing the contributions missed by second
order perturbation theory. In the $\mu_k = 1$ case these contributions
are significant at higher $k$, being greater than 5\% above $k = 10
\,h^{-1} \:\mathrm{Mpc}^{-1}$---for accurate calculations of the
redshift-space power spectrum on small scales a fully non-linear
calculation is essential.

\begin{figure}
\begin{center}
\includegraphics[width=0.8\textwidth]{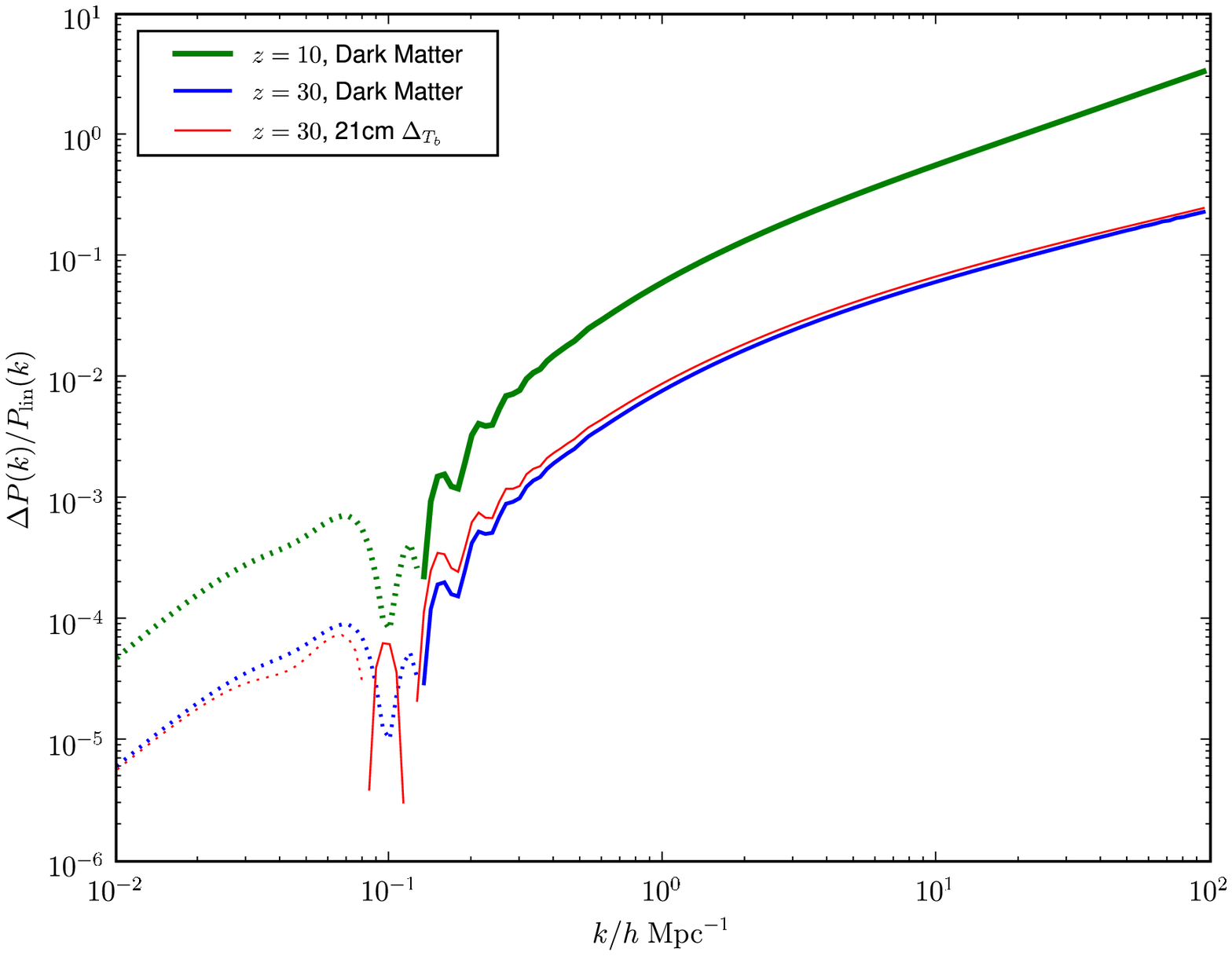}
\end{center}
\caption{The ratio of the non-linear contributions to the linear
  predictions for the dark matter redshift-space power spectrum at
  redshifts of $z = 10$ and $30$, and the 21cm brightness temperature
  power spectrum, all for $\mu_k = 1.0$. In all cases the non-linear
  contributions become significant at high $k$, whilst in the low
  redshift dark matter case they become dominant for $k$ approximately
  greater than $10 \; h \mathrm{Mpc}^{-1}$. This also shows the 21cm
  non-linear corrections are of the same magnitude as the dark matter
  corrections.}
\label{fig:comp_z}
\end{figure}

In Figure \ref{fig:comp_z} the size of the non-linear contributions
from redshift distortions at redshifts of $z = 10$ and $z = 30$ is
compared for dark matter and 21cm brightness temperature
perturbations. On small scales the boosting of power means that
non-linear effects are increasingly important in comparison to the
linear prediction. At a redshift of $z = 10$ their dominance at
reasonable scales means that they are potentially observationally
relevant. This is still true for the 21cm spectra, and we discuss the
consequences of this in Section \ref{sec:21cm}.

\begin{figure}
\begin{center}
\includegraphics[width=0.8\textwidth]{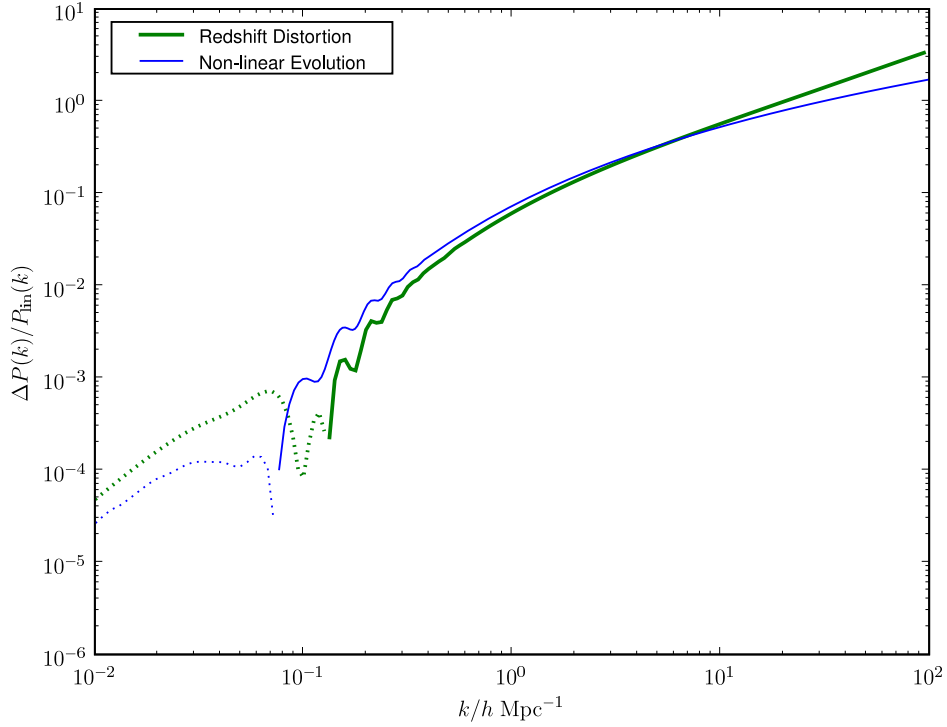}
\end{center}
\caption{The ratio of the corrections due to non-linear redshift-space
  distortions ($\mu_k=1$) and non-linear evolution contributions to
  the linear dark matter power spectrum at a redshift of $z = 10$. The
  redshift distortion corrections are of roughly the same magnitude as
  those from non-linear growth at all scales, and become greater on
  smaller scales--- both effects should be thought of as equally
  important when considering modes not orthogonal to the line of
  sight.}
\label{fig:comp_nl}
\end{figure}

In Figure \ref{fig:comp_nl} the size of the non-linear contributions
from redshift distortions is compared to that from non-linear growth
(calculated using 3rd-order perturbation
theory~\cite{Makino92,Scoccimarro:2004tg}). The contributions are of
equivalent importance at all scales.

\section{Angular Correlations on the Curved Sky}
\label{sec:curvedsky}

\noindent The redshift space power spectrum that we calculated in the
previous section, like the first order result, contains an explicit
anisotropy within the small observed volume due to the direction
defined by the line of sight. Whilst useful for consideration of
localized distortions in redshift space, we should remember that each
observer in the universe should see a statistically isotropic light
cone if the universe is statistically isotropic and homogeneous.  It
is the angular correlation between different redshifts on the light
cone that is directly observable. The most natural descriptions for
the whole sky should take this directly into account, separating out
the radial distances and displacements. In this section we calculate
the angular correlation function $\xi_s(x,y,\mu)$ which correlates
observations at points at redshifts $z$ and $z'$ separated by angle
$\cos^{-1}\mu$; and the angular power spectrum $C_l(z,z')$ giving the
correlation of multipoles $l$ at different redshifts $z$ and $z'$.

Our starting point is to calculate the correlation function between
positions $\vz$ and $\vz'$ in redshift space. This is achieved by
taking the inverse transform of \eqref{eq:dkdq} yielding
\begin{equation}
  \left\langle \Delta_s(\vz)\, \Delta_s(\vz')\right\rangle = \iint \frac{d^3k\, d^3q}{(2\pi)^6} e^{i\left[\vk \cdot \vz + \vq \cdot \vz'\right]} \bigl\langle  \Delta_s(\vk) \, \Delta_s(\vq) \bigr\rangle   \; .
\end{equation}
This is effectively the forward and inverse transform of our starting
point (usually a redundant process), we have required it to eliminate
the unwanted $\kpar$ terms. Substituting \eqref{eq:dkdq} into the
above (with the delta-function that was suppressed from
Eq.~\eqref{eq:deltak_delta}) we have:
\begin{align}
\left\langle \Delta_s(\vz)\, \Delta_s(\vz')\right\rangle = & \int \frac{d^3k\, d^3q\,d^3x\,d^3y}{(2\pi)^6}\; e^{i \left[\vk \cdot \left(\vz - \vx\right) + \vq \cdot \left(\vz' - \vy\right)\right] } \biggl[ e^{-\frac{1}{2} \left[\kpar^2 C_{\phi}(\vx,\vx)  + \qpar^2  C_\phi(\vy,\vy) + 2 \kpar\qpar C_\phi(\vx,\vy)\right]} \notag \\
& \times \Bigl[1 + C_{\Delta_s}(\vx,\vy) - i \qpar C_{\Delta\phi}(\vx,\vy) - i \kpar C_{\Delta\phi}(\vy,\vx) - \kpar \qpar C_{\Delta\phi}(\vx,\vy) C_{\Delta\phi}(\vy,\vx)\Bigr] \biggr] - 1.
\end{align}
The term in the large square brackets above is a function only of
$\kpar = \vk \cdot \vnhat_x$ and $\qpar = \vq \cdot \vnhat_y$, and
thus we can integrate out the perpendicular components of $\vk$ to
give the delta functions $\delta^2(\vz_\perp)$ and
$\delta^2(\vz'_\perp)$. These effectively constrain $\vx$ and $\vy$
enforcing them to be parallel to $\vz$ and $\vz'$ respectively. Given
that redshift distortions displace only along the line of sight this
is what we should expect. This leaves the integral:
\begin{align}
\left\langle \Delta_s(\vz)\, \Delta_s(\vz')\right\rangle = & \int \frac{d\kpar \, d\qpar \, dx \, dy}{(2\pi)^2}\; e^{i \left[\kpar \left(z - x\right) + \qpar \left(z' - y\right)\right] } \biggl[ e^{-\frac{1}{2} \left[\kpar^2 C_{\phi}(\vx,\vx)  + \qpar^2  C_\phi(\vy,\vy) + 2 \kpar\qpar C_\phi(\vx,\vy)\right]} \notag \\
& \times \Bigl[1 + C_{\Delta}(\vx,\vy) - i \qpar C_{\Delta\phi}(\vx,\vy) - i \kpar C_{\Delta\phi}(\vy,\vx) - \kpar \qpar C_{\Delta\phi}(\vx,\vy) C_{\Delta\phi}(\vy,\vx)\Bigr] \biggr] -1,
\end{align}
where now the vectors $\vx = x \vnhat_z$ and $\vy = y
\vnhat_z'$. Conveniently this is now an integral of Gaussian form in
the variables $\kpar$ and $\qpar$ that we can analytically
evaluate. Writing these as the vector $\trans{\vst} = (\kpar, \qpar)$,
we recast the integral as
\begin{align}
\left\langle \Delta_s(\vz)\, \Delta_s(\vz')\right\rangle = & \int \frac{dx \, dy \, d^2\vst}{(2\pi)^2}\; \exp{\left[-\frac{1}{2} \trans{\vst} \ma \vst - i \trans{\vsu}\cdot\vst\right]} \notag \\
& \times \Bigl[1 + C_{\Delta}(\vx,\vy) - i \vst_2 C_{\Delta\phi}(\vx,\vy) - i \vst_1 C_{\Delta\phi}(\vy,\vx) - \vst_1 \vst_2 C_{\Delta\phi}(\vx,\vy) C_{\Delta\phi}(\vy,\vx)\Bigr] -1,
\end{align}
where
\begin{subequations}
\begin{align}
\trans{\vsu} & = \left(x - z,\: y - z'\right) \; ,\\
\ma & = \begin{pmatrix} C_\phi(\vx,\vx) & C_\phi(\vx,\vy) \\ C_\phi(\vx,\vy)
  & C_\phi(\vy,\vy) \end{pmatrix} \; .
\end{align}
\end{subequations}
The prototype for this integral is
\begin{equation}
\int d^2\vst \; \exp{\left(-\frac{1}{2} \trans{\vst} \ma \vst - i \trans{\vsu}\cdot\vst\right)} = 2\pi\: \mathrm{det}^{-1/2}\ma \:\; e^{-\frac{1}{2} \trans{\vsu}\ma^{-1}\vsu} \; .
\end{equation}
Further moments can be generated by taking derivatives with respect to
the vector $\vsu$ as done to construct \eqref{eq:exp_val}. Putting
this together, the correlation function is given by a two dimensional
integral in the radial distances $x$ and $y$,
\begin{multline}
\label{eq:corr_full}
\left\langle \Delta_s(\vz)\, \Delta_s(\vz')\right\rangle =  \int dx \, dy \frac{e^{-\frac{1}{2} \trans{\vsu}\ma^{-1}\vsu}}{2\pi|\ma|^{1/2}} \Bigl[1 + C_{\Delta}(\vx,\vy) - [\ma^{-1}\vsu]_2 C_{\Delta\phi}(\vx,\vy) - [\ma^{-1}\vsu]_1 C_{\Delta\phi}(\vy,\vx) \\
- \left[\ma^{-1}- \ma^{-1}\vsu\trans{\vsu}\ma^{-1}\right]_{12} C_{\Delta\phi}(\vx,\vy)C_{\Delta\phi}(\vy,\vx) \Bigr] -1 .
\end{multline}
The result expresses the redshift-space correlation function roughly
as the integral of the correlations functions against the Gaussian
distribution of the velocities at the two points.

Given the isotropy of the correlation functions $C_a(\vx,\vy)$ they
must depend only on the lengths $x = \left\lvert\vx\right\rvert$, $y =
\left\lvert\vy\right\rvert$ and the angle between them of which we
take the cosine $\mu = \vnhat_z \cdot \vnhat_{z'}$, and so we write
them as $C_a(\vx,\vy) = \xi_a(x,y,\mu)$. Similarly $\left\langle
  \Delta_s(\vz)\, \Delta_s(\vz')\right\rangle$ depends only on $z$,
$z'$ and $\mu$, and we write it as $\xi_s(z,z',\mu)$. So in its final form the correlation function is
\begin{multline}
\xi_s(z,z',\mu) =  \frac{1}{2\pi} \int dx \, dy \;\mathrm{det}^{-1/2}\ma \:\; \exp{\left(-\frac{1}{2} \trans{\vsu}\ma^{-1}\vsu \right)}  \\
\times \biggl[1 + \xi_{\Delta}(x,y,\mu) - [\ma^{-1}\vsu]_2 \xi_{\Delta\phi}(x,y,\mu) - [\ma^{-1}\vsu]_2 \xi_{\Delta\phi}(y,x,\mu) \\
+ \left[\ma^{-1} - \ma^{-1}\vsu\trans{\vsu}\ma^{-1}\right]_{12} \xi_{\Delta\phi}(x,y,\mu) \xi_{\Delta\phi}(y,x,\mu) \biggr] - 1 \; .
\end{multline}
This closed form expression completely describes the non-linear
redshift-space distortions and unlike the flat sky approach we have
yet to make any assumptions about the change along the light
cone. This ensures it is easy to incorporate the evolution of the
fields and the background~\cite{Barkana:2005jr}. A similar result,
specific to the flat-sky was found in \cite{Bharadwaj:2001zf}.

\begin{figure}
\begin{center}
\includegraphics[width=0.8\textwidth]{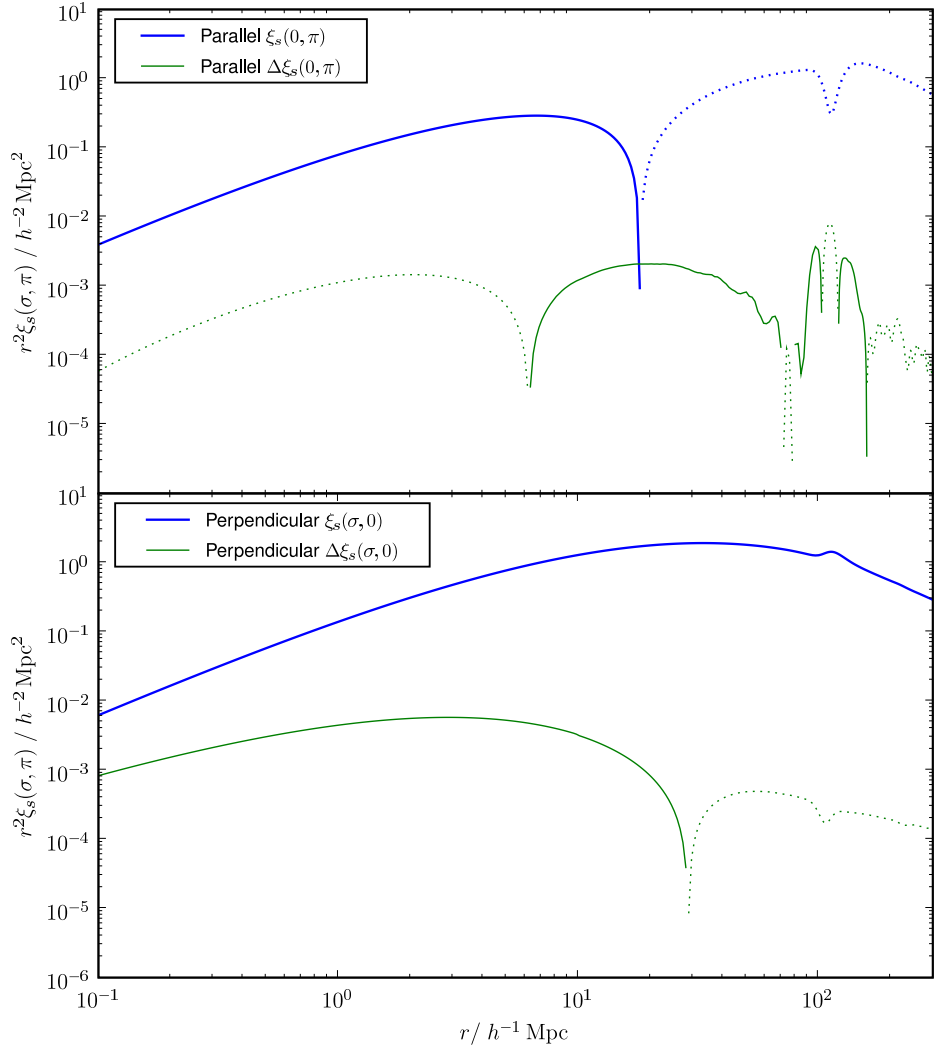}
\end{center}
\caption{The redshift-space correlation function $\xi_s(\sigma,\pi)$
  at a redshift $z = 10$ for Dark Matter. The top panel illustrates
  the full correlation function, $\xi_s(\sigma,\pi)$, and the
  non-linear contributions to it, $\Delta\xi_s(\sigma,\pi)$, in the
  parallel direction. The lower panel the same, but in the
  perpendicular direction. The acoustic peak can clearly be seen at a
  comoving scale of around $100 \; h^{-1}\, \mathrm{Mpc}$. The sharp
  peaking in the non-linear contributions above $10 \; h^{-1}\,
  \mathrm{Mpc}$ is largely due to the smoothing effect on the acoustic
  peak, and small perturbations around the zero crossing points that
  are large relative to the linear result.}
\label{fig:comp_corr}
\end{figure}

The correlation function is frequently used in the study of baryon
acoustic oscillations (BAO) to describe the distortions observed on
small patches of sky. There it is conventionally denoted
$\xi(\sigma,\pi)$, correlating points separated by a comoving distance
along the line-of-sight of $\pi$ and perpendicular to it $\sigma$,
where the curvature of the sky is neglected. This gives a total
separation $r = \sqrt{\pi^2 + \sigma^2}$, and we place the points an
average distance $\bar{z}$ from the origin. We can calculate the
non-linear equivalent in the flat-sky by picking $z$, $z'$ and $\mu$
equivalent to $\sigma$, $\pi$ and $\bar{z}$:
\begin{subequations}
\begin{align}
z   & = \sqrt{1 + (\sigma / 2 \bar{z})^2} \left(\bar{z} + \pi / 2\right) \; ,\\
z'  & = \sqrt{1 + (\sigma / 2 \bar{z})^2} \left(\bar{z} - \pi / 2\right) \; ,\\
\mu & = 2 \tan^{-1}\left(\frac{\sigma}{2 \bar{z}}\right)
\end{align}
\end{subequations}
Figure~\ref{fig:comp_corr} shows $\xi_s(\sigma,\pi)$ and the
difference between the linear and non-linear results,
$\Delta\xi_s(\sigma,\pi) = \xi_s(\sigma,\pi) -
\xi_{\alpha}(\sigma,\pi)$, for the exactly parallel and perpendicular
cases, calculated by the above procedure. We discuss how to calculate
the flat-sky linear correlation function $\xi_\alpha(\sigma, \pi)$ in
Appendix~\ref{app:correlation}. As in the previous cases the
non-linear effects change the correlations on small scales by
significant amounts (around 10\%), though the effect for the parallel
case is much smaller than the perpendicular. In the parallel case
there is a smoothing of the acoustic peak, resulting in a small
suppression of around 3\%.

The distortions introduced on the full sky are perhaps most
conveniently described by the the angular correlation function, giving
the correlation of multipoles on different redshift slices. The $l$-th
multipole moment $C_l(z,z')$ is found by integrating with
$\mathcal{P}_l(\mu)$, the $l$-th Legendre polynomial, that is
\begin{equation}
C_l(z,z') = 2\pi \int d\mu\: \mathcal{P}_l(\mu)\, \xi_s(z,z',\mu) \; .
\end{equation}
Substituting \eqref{eq:corr_full} gives the final integral for the
angular correlation (at $l>0$) for redshifts $z$ and $z'$:
\begin{multline}
\label{eq:curved}
C_l(z,z') =  2\pi \int d\mu \, dx \, dy \; \mathcal{P}_l(\mu)\: \frac{1}{2\pi\,\mathrm{det}^{1/2}\ma} \:\; \exp{\left(-\frac{1}{2} \trans{\vsu}\ma^{-1}\vsu \right)}  \\
\times \biggl[1 + \xi_{\Delta}(x,y,\mu) - [\ma^{-1}\vsu]_2 \xi_{\Delta\phi}(x,y,\mu) - [\ma^{-1}\vsu]_2 \xi_{\Delta\phi}(y,x,\mu) \\
+ \left[\ma^{-1} - \ma^{-1}\vsu\trans{\vsu}\ma^{-1}\right]_{12} \xi_{\Delta\phi}(x,y,\mu) \xi_{\Delta\phi}(y,x,\mu) \biggr] \; .
\end{multline}

In getting to this result we have avoided most of the common
assumptions made when considering redshift-space problems,
non-evolving field statistics and the distant observer approximation
(at least for density fields like the matter perturbation, and source
number counts). This ensures it naturally incorporates any large angle
geometric effects that are not included by taking the flat-sky power
spectrum onto the full sky. For further discussion of this see
Ref.~\cite{Matsubara:1999du,Papai:2008bd}.

The correlation functions $\xi_a(x,y,\mu)$ encapsulate all the
information required to calculate the power spectrum, and our
formulation above remains completely general. To construct the
correlations we must consider several effects, notably the underlying
matter correlations and growth along the light cone. In Appendix
\ref{app:correlation} we consider how to calculate the correlation
functions.

If choosing to use the distant observer approximation, or dealing
approximately with radiative fields such as the brightness
temperature, we can follow through the same analysis above but
starting from the contents of Section~\ref{sec:radiative}. This leads
to the notationally simpler result
\begin{equation}
C_l(z,z') =  2\pi \int d\mu dx \, dy \, \frac{e^{-\frac{1}{2} \trans{\vsu}\ma^{-1}\vsu}}{2\pi\,\mathrm{det}^{1/2}\ma}
\Bigl[\xi_{\alpha}(x,y,\mu)
- \left[\ma^{-1} - \ma^{-1}\vsu\trans{\vsu}\ma^{-1}\right]_{12} \xi_{\alpha\phi}(x,y,\mu)\xi_{\alpha\phi}(y,x,\mu) \Bigr] \mathcal{P}_l(\mu) \; .
\end{equation}
Note that this is exact at lowest order, only dropping ${\cal
  O}(\phi/x)$ curved-sky terms at higher order, provided we use the
correct forms of $\alpha$ for radiative and spatial fields.

\begin{figure}
\begin{center}
\includegraphics[width=0.8\textwidth]{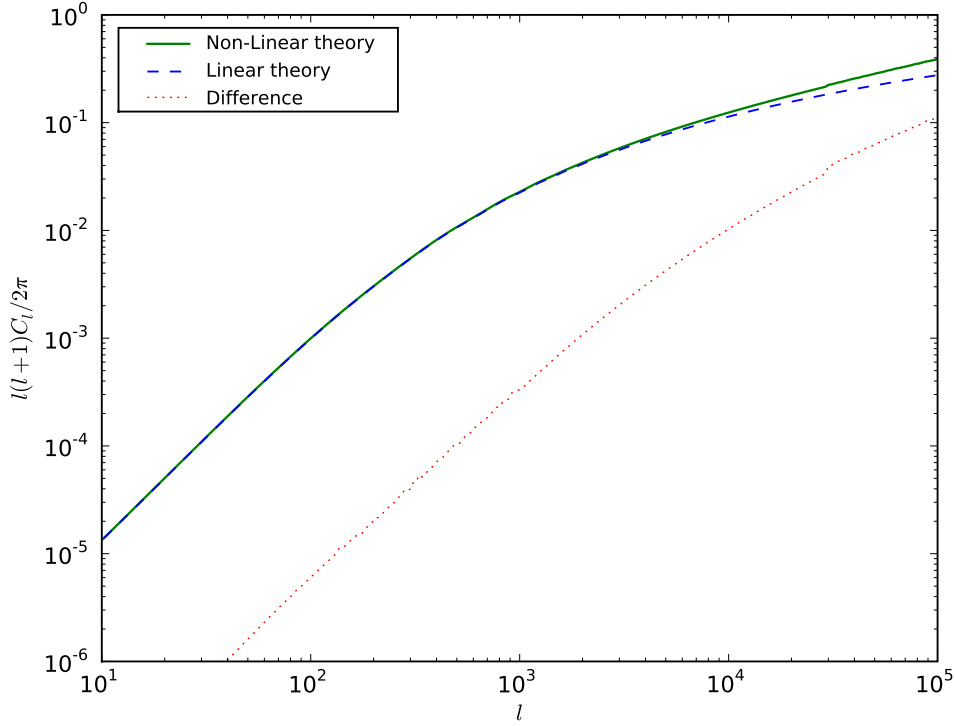}
\end{center}
\caption{The equal redshift dark matter angular power spectrum for $z
  = z' = 10$. We plot the redshift space power spectrum from the
  linear theory prediction, the non-linear result of
  Eq.~\eqref{eq:curved} and the difference between the two. The
  correction is 1\% at $l \approx 520$ and becomes greater than 10\%
  above $l \approx 11000$.}
\label{fig:comp_mult}
\end{figure}

In Figure~\ref{fig:comp_mult} we plot the redshift-space dark matter
power spectrum for slices of zero separation at a redshift of $z =
10$, comparing the fully non-linear result to the linear theory
(described in detail in \cite{Datta:2006vh}). The linear result is
essentially the generalisation of the Kaiser result onto the full sky,
taking the form
\begin{equation}
C_l(z,z') = \frac{2}{\pi} \int_0^\infty dk\,k^2 \Bigl[j_l(k z) j_l(k z') P_\Delta(k) - \bigl[j_l(k z) j''_l(k z') + j''_l(k z) j_l(k z')\bigr] P_{\Delta v}(k) + j''_l(k z) j''_l(k z') P_{v}(k)\Bigr] \; .
\end{equation}
At large $l$ we get a boost in power over the linear-theory results as
we would expect from the previous discussion on the flat sky. The
effect at small $l$ is less than 1 \%, though this is significantly
more than the effect on the power spectrum at equivalent wavenumbers
--- the lack of intrinsic power at large scales means that the large
scale signal in the angular power spectrum is primarily sourced from
much higher wave numbers where the non-linear effects are greater. The
increases on large scales are a consequence of this with a possible
contribution from including the distant observer terms, though we have
not disentangled their relative importance. Figure~\ref{fig:comp_mult}
does not obviously show the acoustic peaks, this is a consequence of
the fact we do not include a window function in $z$ --- the narrow
band tends to smooth out such features.


\newcommand{\tbar}{\overline{T}_b}
\newcommand{\xbar}{\overline{x}_i}

\section{Component Separation for High Redshift 21cm Observation}
\label{sec:21cm}

\noindent The observation of neutral hydrogen through the 21cm
spin-flip transition provides a unique opportunity for probing the
high-redshift Universe. In principle observations can give a
three-dimensional view of structure in the Universe from a redshift of
$z = 300$ all the way down to the epoch of reionization at around $z =
6$ and below. The signal seen in absorption at $z\agt 30$ is expected
to be nearly linear, with significant redshift
distortion~\cite{Bharadwaj:2004nr}, and containing angular structure
down to the baryon pressure-support
scale~\cite{Scott90,Loeb:2003ya,Lewis:2007kz}. With so many modes
cosmology could be constrained to very high precision. Although
nearly-linear, small non-linear effects will still be very important
if observations are to be used reliably, so a non-linear treatment of
redshift-distortions will be essential. At redshifts below $z\alt 30$
the signal is expected to become much more complicated due to the
presence of Lyman-$\alpha$ photons and ionizing sources. Learning
about cosmology from these observations would require detailed
modelling of complicated and poorly understood astrophysics (see
Ref.~\cite{Furlanetto:2006jb} for a review). Likewise source number
counts (in 21cm or otherwise) are hard to model reliably due to scale
and time-dependent bias. However in both cases the velocities are
likely to be much closer to linear theory, making them a much more
robust probe of the underlying cosmological perturbations. If redshift
distortions can be isolated, they therefore represent a powerful way
to learn about cosmological perturbations from present and near-future
observations (e.g. see recent work in
Refs.~\cite{Guzzo:2008ac,Song:2008qt} and references therein).

The quantity we are interested in for 21cm observations is the
brightness temperature $T_b$, with perturbation $\Delta_{T_b}$. In
real space this is given approximately by
\begin{equation}
\label{eq:brighttemp}
\Delta_{T_b} = \beta_b \delta_b + \beta_x \delta_x + \beta_\alpha \delta_\alpha + \beta_{T_K} \delta_{T_K},
\end{equation}
where $\delta_b$ is the baryon perturbation, $\delta_x$ the ionization
fraction perturbation, $\delta_\alpha$ the Lyman-$\alpha$ coupling
perturbation, and $\delta_{T_K}$ the perturbation in the gas kinetic
temperature. The $\beta_i$ depend on the background evolution, for a
more detailed overview see Ref.~\cite{21cm:furlanetto2006}. Note that
throughout this section we return to the flat-sky approximation.

Although the astrophysics that affects the 21cm signal is very
interesting in its own right, to constrain primordial perturbations
more directly we would like to determine of the power spectrum of
matter perturbations $P_\delta(k)$. Unfortunately $\Delta_{T_b}$ mixes
the astrophysical information from the ionization fraction,
Lyman-$\alpha$ coupling and gas temperature in with the cosmological
information we desire. However redshift-space distortions add in
further information directly linked to the matter perturbations in the
approximation in which the source velocities follow the linear CDM
velocity. The linear redshift-space power spectrum can then be written
\begin{equation}
P_{s,T_b}(\vk) = P_{T_b}(k) + 2\mu_k^2 P_{T_b, v}(k) + \mu_k^4 P_v(k)\; ,
\end{equation}
where the $P_{T_b}(k)$ is the power spectrum of brightness temperature
fluctuations in real space encapsulating all the correlations and
cross-correlations of Eq.~\eqref{eq:brighttemp}. The term
$P_{T_b,v}(k)$ gives the cross-correlation with the velocity
perturbation $\delta_v$. At linear order we see that the $\mu_k^4$
contribution is entirely the matter power spectrum, giving a possible
method of separation without needing to understand the detailed
physics encapsulated in $P_{T_b}(k)$ and
$P_{T_b,v}(k)$~\cite{Barkana:2004zy,Mao:2008ug}.  However this approach is
reliant on the use of the linear expansion: as we can see in
Eq.~\eqref{eq:flatsky_final} the full angular behaviour is much more
complicated, and does not lend itself to an easy separation in powers
of $\mu_k$. So we should expect this naive separation method to
perform badly wherever the non-linear contributions are important.

\begin{figure}[htpb]
\begin{center}
\includegraphics[width=0.8\textwidth]{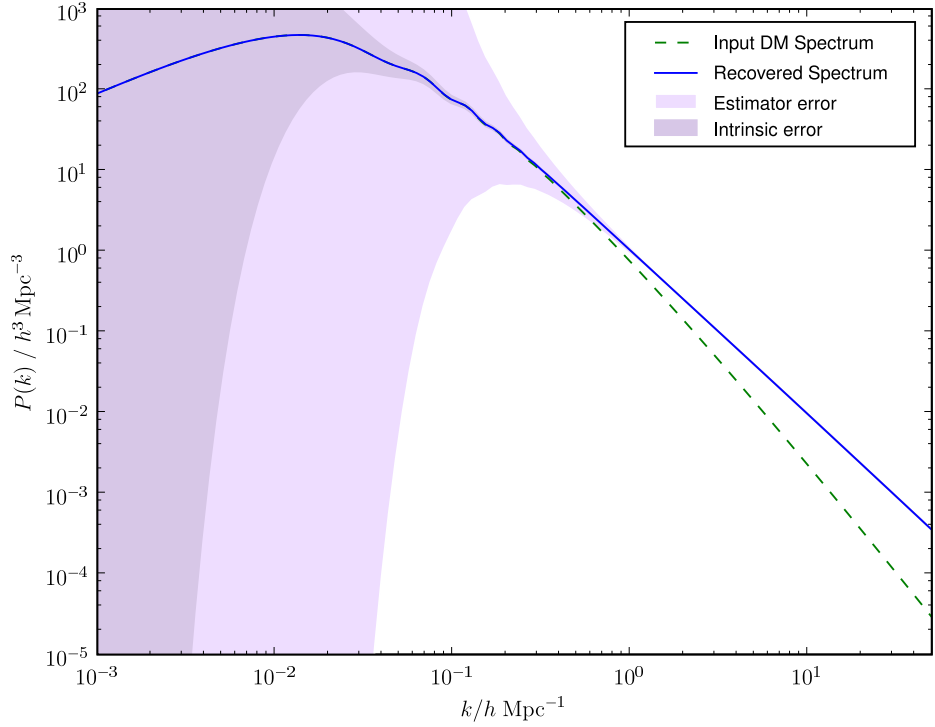}
\end{center}
\caption{The input real-space matter power spectrum at $z = 10$
  compared to that recovered via the estimator $\hat{P}_v$ given by
  Eq.~\eqref{eq:phat}. We include the errors (shading) for a Hubble
  volume sized survey at $z = 10$ assuming a binning of $\Delta k / k
  = 0.1$. The estimator error corresponds to the error if we used the
  estimator $\hat{P}_v$ discussed in the text. We also plot the
  intrinsic error that would be seen if we could measure the modes
  $\delta_v$ directly. At high $k$ the recovered spectrum differs
  dramatically from that input due to the importance of higher-order
  terms, giving a significant systematic bias outside of the
  statistical errors. }
\label{fig:compsep}
\end{figure}

To test this in an ideal case, we calculate the theoretical dark
matter power spectrum in redshift space at a redshift $z = 10$. Taking
100 points equally spaced in $\mu_k$ we integrate with
$\mathcal{P}_4(\mu)$, the fourth Legendre polynomial, to isolate the
$\mu^4_k$ contribution. With the appropriate normalization, our
estimator, exact within linear theory, is
\begin{equation}
\label{eq:phat}
\hat{P}_v(k) = \frac{315}{16} \int^1_{-1} d\mu_k P_s(k, \mu_k) \mathcal{P}_4(\mu_k) \; .
\end{equation}
We compare the underlying power spectrum with that recovered via this
method in Fig.~\ref{fig:compsep}. The recovered power spectrum is
artificially high at large $k$. Repeating this with a power spectrum
generated from the linear result as expected reproduces the input
exactly.
In Appendix~\ref{app:series} we calculate the leading-order non-linear
correction on small scales, which shows that we have a direct
$\mu_k^4$ contribution taking the form $\mu_k^4 P_{v}(k)
\xi_\alpha(0)$. This combines the power spectrum we desire with the
source point variance of large scales, mixing in information from the
large-scale astrophysics, and is a significant contributor to the bias
of this estimator. Correct interpretation of high-redshift
observations on small scales will therefore require a more
sophisticated analysis that accounts for the complicated angular
behaviour introduced at non-linear order, or modelling of the
astrophysics in a realistic and accurate manner.

To assess whether any bias is significant, we can calculate the
variance of this estimator given a few assumptions about the density
of the sampling we can perform in $\vk$-space. We assume a survey of a
large volume of the universe $V$ centered at a redshift $z$, that has
a small angular span such that we are still in the flat sky. We define
an estimator for the power spectrum at a wavenumber $k$, and line of
sight angle $\cos^{-1}\mu$, that using a suitable weighting function
$w_\vk(k, \mu)$ is defined by
\begin{equation}
\hat{P}_s(k, \mu) = \sum_\vk w_\vk(k, \mu) \left\lvert\Delta_\vk\right\rvert^2 \; ,
\end{equation}
where the summation is over all the samples in Fourier space. We are
free to choose any weighting function such that the ensemble average
$\langle \hat{P}_s(k, \mu) \rangle =
P_s(k, \mu)$. Calculating the $\mu$-covariance of this estimator we
find
\begin{equation}
\label{eq:phatkmu}
\Bigl\langle \Delta\hat{P}_s(k, \mu_1) \Delta\hat{P}_s(k, \mu_2) \Bigr\rangle = 2 \sum_\vk w_\vk(k, \mu_1) w_\vk(k, \mu_2) P_s(\vk)^2 \; .
\end{equation}
From \eqref{eq:phat} the variance of the estimator $\hat{P}_v$ is given by
\begin{equation}
\label{eq:varphat}
\Bigl\langle\Delta\hat{P}_v(k)^2\Bigr\rangle = \frac{99225}{256} \int d\mu_1 d\mu_2 \mathcal{P}_4(\mu_1) \mathcal{P}_4(\mu_2) \bigl\langle \Delta\hat{P}_s(k, \mu_1) \Delta\hat{P}_s(k, \mu_2) \bigr\rangle
\end{equation}
Ideally we would optimise the weights $w_\vk(k, \mu)$ to minimize the
variance of $\hat{P}_v$, but for our purposes it will suffice to pick
a representative form--- averaging in bins of width $\Delta k$ and
$\Delta \mu$. This picks out $k^2 \Delta k \Delta \mu V
/ (2\pi)^2 = n(k, \mu) \Delta k \Delta\mu$ modes and we assume that our samples in $\mu$ are spaced
widely enough that the summation of \eqref{eq:phatkmu} contributes
only when $\mu_1$ equals $\mu_2$, giving
\begin{equation}
w_\vk(k, \mu) = \begin{cases} 1 / n(k,\mu)\Delta k\Delta\mu & \left\lvert \vk \right\rvert \in \left[k, k+\Delta k\right], \vnhat\cdot\vk \in \left[\mu, \mu+\Delta\mu\right] \\
0 & \text{otherwise}
\end{cases}
\end{equation}
Given the finite samples in $\mu$ we can draw, we approximate the
integrals of \eqref{eq:varphat} into summations
\begin{equation}
\label{eq:varphat2}
\Bigl\langle\Delta\hat{P}_v(k)^2\Bigr\rangle \approx \frac{99225}{128} \sum_{i j} (\Delta \mu)^2 \mathcal{P}_4(\mu_i) \mathcal{P}_4(\mu_i) \sum_\vk w_\vk(k, \mu_i) w_\vk(k, \mu_j) P_s(\vk)^2 \; .
\end{equation}
Substituting for $w_\vk(k, \mu)$ connects the summations over $i$ and
$j$, and writing the density of modes with a wavevector of length $k$
as $n(k) = 4\pi k^2 V / (2\pi)^3$ we have
\begin{equation}
\label{eq:varphat3}
\Bigl\langle\Delta\hat{P}_v(k)^2\Bigr\rangle \approx \frac{99225}{64} \frac{1}{n(k) \Delta k}\sum_{i} (\Delta \mu) \mathcal{P}_4(\mu_i)^2 P_s(k, \mu_i)^2 \; .
\end{equation}
At low-$k$ the Kaiser result is a reasonable approximation, and thus
we use this to calculate the variance. Taking the continuum limit of
the summation, we can perform the angular integral analytically for
fields with linear bias. The lower bound for the error is the unbiased
tracer $b = 1$ giving the numerical result
\begin{equation}
\frac{\Delta\hat{P}_v(k)}{\hat{P}_v(k)} \approx 50 \frac{1}{\sqrt{n(k) \Delta k}} \; .
\end{equation}
This shows that the errors in calculating the underlying velocity
power spectrum by this component separation are around $35$ times
larger than those we would find if we could directly measure velocity
modes within the observed volume. This increases the lowest $k$ we
could infer by around a factor of $10$. The plot in
Fig.~\ref{fig:compsep} illustrates the dark matter tracing case for
which $b = 1$ and the errors are exact. For 21cm we expect to find a
large bias and thus the errors are dominated from the contribution of
the variance of the $P_\Delta(k)$ term. Asymptotically, for large bias
\begin{equation}
\frac{\Delta\hat{P}_v(k)}{\hat{P}_v(k)} \approx 19 \frac{b^2}{\sqrt{n(k) \Delta k}} \; .
\end{equation}
To overcome this \cite{McDonald:2008sh} suggest that combining
multiple tracers with distinct biases may be able to reduce this error
down closer to the intrinsic level. Though obviously useful for lower
redshift surveys where many indepedent tracers can be found as
different galaxy populations, they suggest it may be possible to use
this for 21cm observations by applying certain non-linear
transformations to the observed field. This method, however, is
dependent upon the linear result being correct, restricting its
applicability to large scales.

One further ramification is that the higher order angular effects from
the non-linear distortions blur any distinction between the
Alcock-Pacy\'{n}ski (AP effect) and those of redshift
distortions. This may produce complications for methods that seek to
obtain cosmological constraints through the AP
effect~\cite{Barkana:2005nr}. Generally these provide constraints by
tuning parameters until angular dependence of $\mu^6$ and above is
eliminated (which is zero for linear redshift-space
distortions). However at large $k$ the non-linearities in redshift
space ensure that even in the correct cosmology, contributions from
higher powers of $\mu$ will be non-zero and tuning them to zero would
be introducing errors in the parameter fitting. This significance of
this is unknown, it may or may not be that the nearly linear low $k$
modes are sufficient to produce constraints unfettered by this.


\section{Conclusions}

\noindent We have shown how to calculate the non-linear effects of
redshift distortion on the power spectrum in the approximation of
Gaussian fields. On small scales the non-linear contributions are
important for modes with a component along the line of sight, even at
high redshift. Superposition of small-scale power on larger-scale
linear modes gives a boost in power on small scales comparable to that
from non-linear structure growth. On larger scales smearing by small
scale velocities leads to a suppression of power. Any future attempt
to extract precision cosmology from high-redshift observations will
need to account carefully for these effects. In order to suitably
describe the behaviour on the full-sky we also extended our technique
to allow calculation of the angular correlation function and power
spectrum. These both have the advantage of naturally incorporating
evolution effects of the background and the fields involved, provided
they remain Gaussian.

For a fully consistent analysis the non-linear growth and
non-Gaussianity should also be accounted for though at present our
work does not yet allow this. Despite this we have already
demonstrated that just the non-linearities introduced by the mapping
from redshift to real space significantly complicate any plan to make
accurate measurements of cosmological perturbations by looking for the
angular structure in the redshift-space signal from our light cone.

In addition to having a significant effect on the power spectrum and
correlation function as discussed in this paper, redshift distortions
will also introduce non-Gaussianity.  For example there is a non-zero
bispectrum for modes that are not all orthogonal to the line of sight.
In the approximation of underlying Gaussian fields the method
developed in this paper extends straightforwardly to higher n-point
functions. This signal will have to be accounted for at high accuracy
(along with the bispectrum introduced by non-linear growth) when
attempting to use future high-redshift observations to constrain
primordial non-Gaussianity~\cite{Cooray:2006km,Pillepich:2006fj}.

Our work could also be extended to include lensing, which in the
Gaussian approximation is just another correlated random field that
perturbs points orthogonal to the line of sight.

\section{Acknowledgements}

\noindent JRS acknowledges receipt of an STFC studentship. AL
acknowledges a PPARC/STFC Advanced Fellowship. We thank Anthony
Challinor for useful discussion and suggestions.

\appendix

\section{Evaluating the Correlation Functions}

\label{app:correlation}
\noindent To calculate the redshift-space power spectrum we must be
able to compute the correlation functions $\xi_{\Delta}$,
$\xi_{\Delta\phi}$ and $\xi_\phi$ in terms of the matter power
spectrum. To start, we note that the 3d-Fourier transform of a
radially symmetric function can be simplified dramatically to a 1d
transform
\begin{equation}
\label{eq:fourier_radial}
\int \frac{d^3k}{(2\pi)^3} e^{i\vk\cdot\vr} f(k) \equiv \frac{1}{2 \pi^2} \int^{\infty}_0\! dk\, j_0(k r)\, \bigl[k^2 f(k)\bigr],
\end{equation}
where $j_0(x) = \sin{x} / x$ is the zeroth spherical Bessel
function. We can generalize this to encapsulate the integrals we will
require later on. Expanding in terms of Spherical Harmonics we use the
identities for $e^{i \vk\cdot \vr}$, and $(\vnhat\cdot\vkhat)^n$
\begin{subequations}
\begin{align}
e^{i \vk\cdot \vr} & = \sum_{l m} i^l j_l(kr) Y^*_{lm}(\vkhat) Y_{lm}(\vrhat) \; ,\\
(\vnhat\cdot\vkhat)^n & = 4\pi \sum_{l m} \frac{n!}{(n-l)!!(n+l+1)!!} Y^*_{lm}(\vnhat) Y_{lm}(\vkhat) \; ,
\end{align}
\end{subequations}
where $\vnhat$ is a direction of our choosing. With these we can
easily evaluate integrals of the form
\begin{equation}
\label{eq:fourier_partial}
\int \frac{d^3k}{(2\pi)^3} e^{i\vk\cdot\vr} (\vkhat \cdot \vnhat)^n f(k) = \frac{1}{2 \pi^2} \sum_{l \le n}  i^l \frac{(2l+1) n!}{(n-l)!!(n+l+1)!!} \mathcal{P}_l(\vnhat\cdot\vrhat) \int^{\infty}_0\! dk\,  \, \bigl[k^2 f(k)\bigr] j_l(k r)  \; ,
\end{equation}
where we have used the orthogonality and addition relations of the
Spherical Harmonics. For any $n$ the summation only has non-zero
elements as far as $l = n$, this means for the small $n$ we are
considering the summations will be limited to only a few terms.

Our first assumption is that $\Delta$ is a statistically isotropic and
homogenous scalar (for example the density perturbation). Secondly we
stay with the definition of $\delta_v$ from
Eq.~\eqref{eq:def_delv}. As a reminder, in real space this relates
$\phi$ and $\delta_v$ via
\begin{equation}
\grad\cdot\vv(\vx) = -\mathcal{H} \delta_v(\vx) \; ,
\end{equation}
where $\del^{-2}$ is the inverse Laplacian operator. For observations
tracing the underlying matter distribution, $\delta_v = f \delta_m$
exactly in the pressureless limit. We will use the Fourier space equivalent
\begin{equation}
\vv(\vk) = i \mathcal{H} \frac{\vk}{k^2} \delta_v(\vk) \; .
\end{equation}
We will eventually express the correlations in terms of transforms of
the power spectra defined by
\begin{subequations}
\begin{align}
\left\langle \Delta(\vk; z_x) \Delta(\vq; z_y) \right\rangle & = (2\pi)^3 \delta^3(\vk+\vq) P_\Delta(k; z_x, z_y) \; ,\\
\left\langle \Delta(\vk; z_x) \delta_v(\vq; z_y) \right\rangle & = (2\pi)^3 \delta^3(\vk+\vq) P_{\Delta v}(k; z_x, z_y) \; ,\\
\left\langle \delta_v(\vk; z_x) \delta_v(\vq; z_y) \right\rangle & = (2\pi)^3 \delta^3(\vk+\vq) P_v(k; z_x, z_y) \; ,
\end{align}
\end{subequations}
which correlate Fourier modes at different epochs given by the
redshifts $z_x$ and $z_y$. In linear theory we can write these in
terms of the transfer functions $T$ and the primordial power spectrum
$P_\chi$
\begin{subequations}
\begin{align}
 P_{\Delta}(k;z_x,z_y) & = T_\Delta(z_x,k) T_\Delta(z_y,k) P_\chi(k) \; , \\
 P_{\Delta v}(k;z_x,z_y) & = T_\Delta(z_x,k) T_v(z_y,k) P_\chi(k) \; , \\
 P_v(k;z_x,z_y) & = T_v(z_x,k) T_v(z_y,k) P_\chi(k) \; .
\end{align}
\end{subequations}
Numerical calculation of the power spectra can be done via codes such
as CAMB \cite{Lewis:1999bs}, or for 21cm perturbations CAMB Sources
\cite{Lewis:2007kz}.

The correlation functions can be written in terms of the
correlations of $\Delta$ and $\delta_v$. Denoting $\theta(\vx) =
\del^{-2}\delta_v(\vx)$ for brevity, they are
\begin{subequations}
\begin{align}
C_\Delta(\vx, \vy) & = \left\langle\Delta(\vx) \Delta(\vy) \right\rangle \; ,\\
C_{\Delta\phi}(\vx, \vy) & =  \yhat_i \left\langle\Delta(\vx) v_i(\vy) \right\rangle \; ,\\
C_\phi(\vx, \vy) & = \xhat_i \yhat_j \left\langle v_i(\vx) v_j(\vy) \right\rangle \; .
\end{align}
\end{subequations}
This reduces the problem down to calculating $\langle\Delta(\vx) v_i(\vy)
\rangle$ and $\langle v_i(\vx) v_j(\vy) \rangle$. Given the
statistical homogeneity and isotropy, these can be decomposed into an
isotropic function of the separation $r = \lvert\vx-\vy\rvert$
combined with the admissible angular factors constructed from
$\vrhat$.
\begin{subequations}
\begin{align}
\left\langle\Delta(\vx) \Delta(\vy) \right\rangle & = A(r) \\
\left\langle\Delta(\vx) v_i(\vy) \right\rangle & = \mathcal{H} \, B(r)\, \rhat_i \\
\left\langle v_i(\vx) v_j(\vy) \right\rangle & = \mathcal{H}^2 \,\left[C(r) \,\delta_{ij} + D(r)\, \rhat_{\langle i}\rhat_{j\rangle} \right]
\end{align}
\end{subequations}
where we add the factors of $\mathcal{H}$ for later
convenience. $\langle\Delta(\vx) \Delta(\vy) \rangle$ is scalar
function and is simply the transform of the power spectrum $P_\Delta$
\begin{equation}
A(r) = \frac{1}{2 \pi^2} \int^{\infty}_0\! dk\, j_0(k r)\, k^2 P_\Delta(k; z_x,z_y) \; ,
\end{equation}
where we leave the $z_x,z_y$ dependence implicit.
The other correlation functions are more complicated. There is only
one possible direction the vector correlation function
$\langle\Delta(\vx) v_i(\vy) \rangle$ can lie along, the separation
vector $\vr$. Multiplying by another $\rhat_j$ and contracting, we
explicitly find $B(r)$ by substituting substituting the Fourier
transform and relating this to the cross power spectrum of $\Delta$
and $\delta_v$
\begin{align}
B(r) & = \left\langle\Delta(\vx) \vrhat\cdot\vv(\vy) \right\rangle / \mathcal{H} \notag \\
& = \int \frac{d^3k}{(2\pi)^3}\, e^{i\vk\cdot\vr}\, \frac{i \vk\cdot\vrhat}{k^2}\, P_{\Delta v}(k; z_x, z_y) \notag \\
& = - \frac{1}{2\pi^2} \int_0^\infty dk j_1(kr) k P_{\Delta v}(k; z_x, z_y) \; .
\end{align}
Then correlation of $\langle v_i(\vx) v_j(\vy) \rangle$ forms a rank-2
tensor that we separate into an isotropic part $C(r)$ and the
traceless outer product of $\rhat_i$ and $\rhat_j$ given by
$D(r)$. Taking the trace isolates $C(r)$ and along the same lines as
above we find
\begin{align}
C(r) & = \frac{1}{3} \left\langle\vv(\vx) \cdot\vv(\vy) \right\rangle / \mathcal{H}^2 \notag \\
& = \frac{1}{3} \int \frac{d^3k}{(2\pi)^3}\, e^{i\vk\cdot\vr}\, \frac{1}{k^2}\, P_{v}(k; z_x, z_y) \notag \\
& = \frac{1}{3} \frac{1}{2\pi^2} \int_0^\infty dk j_0(kr) P_{v}(k; z_x, z_y) \; .
\end{align}
Finally we calculate the traceless part $D(r)$
\begin{align}
D(r) & = \frac{3}{2} \left\langle v_i(\vx) v_j(\vy)  \right\rangle (\rhat_i \rhat_j - \frac{1}{3}\delta_{ij}) / \mathcal{H}^2 \notag \\
& = \frac{3}{2} \int \frac{d^3k}{(2\pi)^3}\, e^{i\vk\cdot\vr}\, \frac{1}{k^2}\, P_{v}(k; z_x, z_y) \left[(\vkhat\cdot\vrhat)^2 - \frac{1}{3}\right] \notag \\
& = - \frac{1}{2\pi^2} \int_0^\infty dk j_2(kr) P_{v}(k; z_x, z_y) \; .
\end{align}
With these functions calculated we can now express the correlation
functions in terms of them
\begin{subequations}
\begin{align}
C_\Delta(\vx, \vy) & = A(r) \; ,\\
C_{\Delta\phi}(\vx, \vy) & = \mu_y B(r) \; , \\
C_{\phi}(\vx, \vy) & = \left[C(r) \mu_{xy} + D(r) (\mu_x \mu_y - \frac{1}{3}\mu_{xy})\right] \; .
\end{align}
\end{subequations}

These results are general, to neaten up the notation somewhat we
specialize them to the flat and curved sky cases we have
considered. For the flat-sky $\vxhat = \vyhat = \vnhat$, and so
$\mu_x = \mu_y = \mu_r$ and $\mu_{xy} = 1$. Evolution along the light
cone is also neglected so we evaluate the power spectra at a single
fixed redshift $z$ giving
\begin{subequations}
\begin{align}
\xi_\Delta(r) & = A(r) \; ,\\
\xi_{\Delta\phi}(r, \mu_r) & = \mu_r B(r) \; , \\
\xi_{\phi}(r, \mu_r) & = \left[C(r) - \frac{1}{3} D(r)\right] + \mu_r^2 D(r) \; .
\end{align}
\end{subequations}

For the curved sky, the correlation function is dependent only on the
radial distances of the points, and the angular separation about the
origin $\mu = \mu_{xy}$. In terms of these $r = \left(x^2 + y^2 - 2 x
  y \mu \right)^{1/2}$, $\mu_x = (y \mu - x) /r$ and $\mu_y = (y - x
\mu) /r$ leaving
\begin{subequations}
\begin{align}
\xi_\Delta(x, y, \mu) & = A(r) \; ,\\
\xi_{\Delta\phi}(x, y, \mu) & = \mu_y B(r) \; , \\
\xi_{\phi}(x, y, \mu) & = \mu \left[C(r) -\frac{1}{3} D(r)\right] + \mu_x \mu_yD(r)  \;.
\end{align}
\end{subequations}

In order to calculate the flat-sky linear redshift correlation
function $\xi_\alpha(r,\mu_r)$, we transform the linear redshift-space
power spectrum $P_\alpha(\vk)$, where as we defined earlier $\alpha =
\Delta -\phi'$, the linear perturbation in redshift
space. Transforming Eq.~\eqref{eq:pslin} term by term, again using
Eq.~\eqref{eq:fourier_partial}, we end up with the following
\begin{align}
\xi_{\alpha}(r, \mu_r) = \left[\xi_\Delta^{(0)}(r) + \frac{2}{3}\xi_{\Delta v}^{(0)}(r) + \frac{1}{5} \xi_v^{(0)}(r)\right] - \left[\frac{4}{3}\xi_{\Delta v}^{(2)}(r) + \frac{4}{7} \xi_v^{(2)}(r)\right]\mathcal{P}_2(\mu_r) + \frac{8}{35} \xi_v^{(4)}(r) \mathcal{P}_4(\mu_r) \; ,
\end{align}
where we have defined the correlation-like functions $\xi_a^{(n)}(r)$
by
\begin{equation}
  \xi_a^{(n)}(r) = \frac{1}{2\pi^2} \int_0^\infty dk \:k^2 P_a(k) j_n(k r) \; .
\end{equation}
To use the standard form $\xi_\alpha(\sigma, \pi)$, we simply set $r
= \sqrt{\sigma^2 + \pi^2}$ and $\mu_r = \pi / r$.

\section{Perturbative Series Expansion}
\label{app:perturb}
In this Appendix we discuss the perturbative expansion of
Eq.\eqref{eq:jac_trans}:
\begin{equation}
\label{eq:dzimpl}
\Delta_s(\vs) = \frac{\Delta(\vx)-\phi'(\vx)}{1+\phi'(\vx)} \; , 
\end{equation}
where $\vx= \vs - \phi(\vx)$. This equation is exact for radiative
fields but uses the distant-observer approximation for number
counts. To solve this implicit equation for $\Delta_s$ we turn to the
Lagrange Reversion Theorem\footnote{See
  e.g. \url{http://en.wikipedia.org/wiki/Lagrange_reversion_theorem}}
that will give us the result in terms of a series expansion. The
theorem states that if we have an implicit definition for $v = x + y
f(v)$ then the function $g(v)$ is given by the series
\begin{equation}
g(v)=g(x)+\sum_{k=1}^\infty\frac{y^k}{k!} \frac{\partial^{k-1}}{\partial x^{k-1}}\left(f(x)^kg'(x)\right) \; .
\end{equation}
To obtain $\Delta_s(\vs)$ we make the obvious assignments to obtain
\begin{equation}
\Delta_s(\vs) = \left.\frac{\Delta-\phi'}{1+\phi'}\right|_\vs +  \left.\sum_{k=1}^\infty\frac{1}{k!}\frac{\partial^{k-1}}{\partial \chi^{k-1}} \left[ (-\phi)^k \frac{\partial}{\partial\chi}\left( \frac{\Delta-\phi'}{1+\phi'}\right)\right]\right|_{\vs} \; .
\end{equation}
Expanding $(1+\phi')^{-1} = \sum_m (-\phi')^m$ and grouping terms of
order $n+1$ this simplifies to
\begin{equation}
\Delta_s(\vs) = \left.\sum_{n=0}^\infty \frac{(-1)^n}{n!}\frac{\partial^n}{\partial\chi^n}\left[ (\Delta-\phi')\phi^{n}\right]\right|_{\vs} \; .
\label{reversion_form}
\end{equation}
Perturbative results can be obtained using this series expansion,
though the perturbative result for the power spectrum is actually
obtained more straightforwardly by expansion of the non-perturbative
result as we show in Appendix~\ref{app:series}.  The series result can
also be written with un-grouped terms as
\begin{equation}
\label{series_ungrouped}
1+\Delta_s(\vs) = \left.\sum_{n=0}^\infty \frac{(-1)^n}{n!}\frac{\partial^n}{\partial\chi^n}\left[ (1+\Delta)\phi^{n}\right]\right|_{\vs}.
\end{equation}
Fourier-transforming Eq.~\eqref{reversion_form} we have
\begin{eqnarray}
\Delta_s(\vk) &=& \sum_{n=0}^\infty \frac{1}{n!}\int d^3 \vs\, e^{-i\vk\cdot \vs} \left. \frac{\partial^n}{\partial\chi^n}\left[ (\Delta-\phi')(-\phi)^{n}\right]\right|_{\vs}\\
&\approx& \int d^3x \: e^{-i \vk \cdot \vx} [\Delta(\vx)-\phi'(\vx)] e^{-i \kpar \phi(\vx)},
\end{eqnarray}
which recovers Eq.~\eqref{Deltas_21cm} of the main text. In the second
line we dropped curved-sky corrections from the radial derivatives of
$x^2$ that arise when integrating by parts, which is consistent at
linear but not at higher order.

\section{Perturbative Result for the Redshift Space Power Spectrum}
\label{app:series}

\subsection{General Expansion}
\noindent Given that we have a general method for calculating the full
non-linear result, a perturbative result is perhaps a little crude,
however it provides some insight into the source of the most important
non-linear effects. We develop the perturbation series from our result
for the flat-sky spectrum in terms of the first order source $\alpha$,
\begin{equation}
P_s(\vk) = \int \! d^3r \: e^{-i \vk\cdot\vr}
  \Bigl[\xi_{\alpha}(\vr) - \kpar^2 \xi_{\alpha\phi}(\vr)^2 \Bigr]e^{-\kpar^2 \left[\xi_{\phi}(0) - \xi_{\phi}(\vr)\right]} \; .
\end{equation}
First we expand the exponential
\begin{equation}
P_s(\vk) =  \int \! d^3r \: e^{-i \vk\cdot\vr}
  \Bigl[\xi_{\alpha}(\vr) - \kpar^2 \xi_{\alpha\phi}(\vr)^2 \Bigr] \: \sum_n\frac{1}{n!}\kpar^{2n}\bigl(\xi_{\phi}(\vr) - \xi_{\phi}(0)\bigr)^n \; ,
\end{equation}
and then we re-sum the term in $\xi_\alpha$ such that each term in the
overall summation contains contributions from the same order in the
correlation functions
\begin{equation}
P_s(\vk) = P_\alpha(\vk) + \sum_n \frac{\kpar^{2\left(n+1\right)}}{\left(n+1\right)!} \int \! d^3r \: e^{-i \vk\cdot\vr}
  \Bigl[\xi_{\alpha}(\vr) \xi_{\phi}(\vr) - \xi_{\alpha}(\vr) \xi_{\phi}(0) - \left(n+1\right) \xi_{\alpha\phi}(\vr)^2 \Bigr] \Bigl(\xi_{\phi}(\vr) - \xi_{\phi}(0)\Bigr)^n \; .
\end{equation}
The power spectrum $P_\alpha$ can be written in terms of the power
spectra of $\Delta$ and $\delta_v$, and similarly for the power
spectra of $P_{\alpha\phi}$ and $P_\phi$:
\begin{subequations}
\begin{align}
P_{\alpha}(\vk) & = P_{\Delta}(k) + 2 \mu_k^2 P_{\Delta v}(k) + \mu_k^4 P_{v}(k) \; ,\\
P_{\alpha\phi}(\vk) & = -i \frac{\mu_k}{k} \left[ P_{\Delta v}(k) + \mu_k^2 P_{v}(k) \right] \; ,\\
P_{\phi}(\vk) & = \frac{\mu_k^2}{k^2} P_v(k) \; .
\end{align}
\end{subequations}
Using the convolution theorem we turn the Fourier transform of the
products of correlations into a convolution of the corresponding power
spectra, giving
\begin{multline}
\label{eq:pert_result}
P_s(\vk) = P_\alpha(\vk) + \sum_n \frac{\kpar^{2\left(n+1\right)}}{\left(n+1\right)!} (2\pi)^3 \int \! \dkp[0] \dkp[1] \dkp[2]\: \delta^3(\vk_0 + \vk_1 + \vk_2 - \vk)\, F_n(\vk_2) \\
\times \Bigl[P_{\alpha}(\vk_0) P_{\phi}(\vk_1) - P_{\alpha}(\vk_0) \xi_{\phi}(0) (2\pi)^3 \delta^3(\vk_1)- \left(n+1\right) P_{\alpha\phi}(\vk_0) P_{\alpha\phi}(\vk_1) \Bigr]  \; ,
\end{multline}
where $\xi_\phi(0)$ is the mean squared line of sight velocity at a
point $\xi_\phi(0) = \frac{1}{3}\frac{1}{\clh^2} \la \vv^2 \ra$. The
convolution kernel $F_n(\vk)$ is defined as an $n$-fold convolution of
$P_\phi(\vk) - (2\pi)^3 \xi_\phi(0) \delta^3(\vk)$,
\begin{equation}
F_n(\vk) = (2\pi)^3 \int \dqp[1]\dotsm \dqp[n] \left[P_\phi(\vq_1) - (2\pi)^3 \xi_\phi(0)\, \delta^3(\vq_1)\right] \dotsm \left[P_\phi(\vq_n) - (2\pi)^3 \xi_\phi(0)\, \delta^3(\vq_n)\right] \delta^3(\vq_1 + \dotsb +\vq_n - \vk) \; ,
\end{equation}
or equivalently the Fourier transform of the $n$-th power of $\xi_{\phi}(\vr) - \xi_\phi(0)$:
\begin{equation}
F_n(\vk) = \int d^3r \: e^{-i \vk\cdot\vr}  \Bigl(\xi_{\phi}(\vr) - \xi_{\phi}(0)\Bigr)^n\; .
\end{equation}
\subsection{Second Order Power Spectrum and Asymptotic Behaviour}

In order to gain some intuition into the non-linear redshift space
distortions, we turn to the leading order corrections to the linear
theory. Using Eq.~\eqref{eq:pert_result} we generate the perturbative
results to second order in the power spectrum. The lowest order term
is simply
\begin{equation}
\presup{(1)}P(\vk) = P_{\alpha}(\vk) \; ,
\end{equation}
the linear redshift space power spectrum that we expect. The terms at
the next order are
\begin{equation}
  \presup{(2)}P(\vk) = \kpar^2 \biggl[ -P_{\alpha}(\vk) \xi_\phi(0) + (2\pi)^3\int \dkp[0]\dkp[1] \bigl[P_{\alpha}(\vk_0) P_{\phi}(\vk_1) - P_{\alpha\phi}(\vk_0) P_{\alpha\phi}(\vk_1) \bigr] \delta^3(\vk_0 + \vk_1 - \vk) \biggr] \; ,
\end{equation}
where at second order in our expansion $n = 0$ and $F_n(\vk) =
(2\pi)^3 \delta^3(\vk)$ giving the above. Specializing to the case of
the matter power spectrum $\Delta = \delta$, and expanding out in full
our result is in agreement with that of Ref.~\cite{Heavens:1998es}
when other non-linear effects are neglected.

To investigate the asymptotic behaviour as $\vk$ becomes large
compared to the turnover in the power spectrum we Taylor expand the
above in this limit. We must be careful to include the contributions
from where either$|\vk_0|$ or $|\vk_1|$ are small, as we expect the
integral to be dominated by contributions from around the turnover. In
this series expansion the leading order terms in $\xi_\phi(0)$ cancel,
leaving the dominant term
\begin{equation}
\presup{(1)}P(\vk) = \kpar^2 \int \frac{d^3\vq}{(2\pi)^3} \biggl[P_\phi(\vk) P_\alpha(\vq) + \frac{1}{2} q^a q^b \Bigl(P_{\phi}(\vq) \left[\del_a \del_b P_{\alpha}(\vk)\right] + P_{\alpha}(\vq) \left[\del_a \del_b P_{\phi}(\vk)\right] \Bigr) - 2 P_{\alpha\phi}(\vq) \left[\del_a P_{\alpha\phi}(\vk)\right] q^a \biggr] \; .
\end{equation}
The $P_{\alpha}(\vq) \left[\del_a \del_b P_{\phi}(\vk)\right]$ term
above is suppressed by a factor of $(q / k)^2$ relative to the other
terms and so we will drop it from our expansion. Averaging out the
angular components of the $\vq$ integrals removes the summations over
$a$ and $b$ and instead directly connects the $\vk$ derivatives with
the line of sight direction, giving
\begin{equation}
\label{eq:smallscale}
\presup{(2)}P(\vk) \approx \kpar^2\biggl[
P_{\phi}(\vk) \int \dqp  P_{\alpha}(\vq)
+2 i\vnhat\cdot\grad_k P_{\alpha\phi}(\vk) \int \dqp P_{\alpha\phi'}(\vq)
+\frac{1}{6} \left[ \grad_k^2 + 2(\vnhat\cdot\grad_k)^2\right]P_{\alpha}(\vk)  \int \dqp P_{\phi'}(\vq)
\biggr] \; .
\end{equation}
Each term is of the form of the power spectrum at $\vk$
(+derivatives) multiplied by a point variance coming from larger
scales. For example the first term gives
\begin{equation}
\kpar^2  P_{\phi}(\vk) \xi_\alpha(0) = \mu_k^4 P_{v}(k) \xi_\alpha(0),
\end{equation}
where the point variance of the first order source is
\begin{equation}
\xi_\alpha(0) = \int \dqp \left[ P_\Delta(q) + \frac{2}{3} P_{\Delta v}(q) + \frac{1}{5} P_v(q)\right].
\end{equation}
The other terms are more complicated, and for the approximation to
make sense the integral ranges should be restricted to scales with
$|\vq| < |\vk|$.  The boost in power on small scales can therefore be
thought of as due to the superposition of sources at that scale
superimposed on large scale linear modes. There are terms up to the
sixth power of $\mu_k$. 

The behaviour on large scales again can be understood by examining the
behaviour for $k \ll k_0, k_1$.  Expanding the integral for small $k$
we have
\begin{equation}
\label{eq:largescale}
\presup{(2)}P(\vk) \approx \kpar^2\left(\frac{1}{3}\int \dqp \frac{1}{q^2} \left[ P_{\Delta}(q)P_{v}(q) -  P_{\Delta v}(q)^2\right] - P_{\alpha}(\vk) \xi_\phi(0)  + \dotsb \right)\; .
\end{equation}
The first term vanishes in the case of perfect correlation between the
source and the velocities, as is the case with one mode of linear
perturbations. In this case the dominant contribution is the
suppression due to the point line-of-sight velocity variance coming
from smaller scales (given by $\xi_\phi(0)$). In the case where the
source and velocities do not correlate on large scales, the integral
is non-zero and positive (by the Cauchy-Schwarz inequality), reducing
the level of suppression.


\bibliographystyle{arxiv}
\bibliography{redshift,antony}

\end{document}